\documentclass[11pt]{article}

\usepackage[a4paper,text={16.8cm,22.4cm}]{geometry}
\usepackage{amsmath,amsfonts,slashed,amssymb,tikz,bm,psfrag,graphicx,xcolor,dsfont}
\usepackage{multicol}
\usepackage{mathrsfs}
\usepackage{array}
\usepackage{multirow}

\RequirePackage[sort&compress,square,comma,numbers]{natbib}
\allowdisplaybreaks
\begin{document}

\begin{titlepage}

\begin{flushright}
\end{flushright}

\vspace{0.1cm}
\begin{center}
\Large\bf
  $\Lambda_b \to p, N^\ast(1535)$ Form Factors from QCD Light-Cone Sum Rules
\end{center}

\vspace{0.5cm}
\begin{center}
{\bf Ke-Sheng Huang$^a$, Wei Liu$^a$, Yue-Long Shen$^b$\footnote{Email: shenylmeteor@ouc.edu.cn, corresponding author}, Fu-Sheng Yu$^{a,c,d}$}\footnote{Email: yufsh@lzu.edu.cn, corresponding author} \\
\vspace{0.7cm}
{\sl
${}^a$\, School of Nuclear Science and Technology,  Lanzhou University, Lanzhou 730000, China}\\
{\sl
${}^{b}$\, College of Physics and Optoelectronic engineering,
Ocean University of China, Qingdao 266100, China}\\
{\sl
	${}^c$\, Frontiers Science Center for Rare Isotopes, and Lanzhou Center for Theoretical Physics, and Key Laboratory of Theoretical Physics of Gansu Province, Lanzhou University, Lanzhou 730000, China}\\
{\sl
	${}^d$\, Center for High Energy Physics, Peking University, Beijing 100871,  China}\\
\end{center}

\vspace{0.2cm}
\begin{abstract}
In this work, we calculate the transition form factors of $\Lambda_b$ decaying into proton and $N^*(1535)$  ($J^P={1/2}^+$ and ${1/2}^-$ respectively) within the framework of light-cone sum rules with the distribution amplitudes (DAs) of $\Lambda_b$-baryon. In the hadronic representation of the correlation function, we have isolated both the proton and the $N^*(1535)$ states so that the  $\Lambda_b \rightarrow p, N^*(1535)$ form factors can be evaluated simultaneously. Due to the less known properties of the baryons, we investigate three interpolating currents of the light baryons and five parametrization models for DAs of $\Lambda_b $. Numerically, our predictions on the $\Lambda_b \rightarrow p$ form factors and the branching fractions of $\Lambda_b\to p\ell \nu$ from the Ioffe or the tensor currents are consistent with the Lattice simulation and the results from the light-baryon sum rules, as well as the experimental data of $Br(\Lambda_b^0\to p\mu^-\bar\nu)$.  The predictions on the form factors of $\Lambda_b \rightarrow N^*(1535)$ are very sensitive to the choice of the interpolating currents so that the relevant measurement could be helpful to clarify the properties of baryons.

\end{abstract}
\vfil
\end{titlepage}

\section{Introduction}

Heavy hadron decays provide an ideal platform for precisely extracting the Cabibbo-Kobayashi-Maskawa (CKM) matrix matrix elements, understanding the QCD dynamics in the heavy hadron system, investigating the standard model (SM) description of the CP violation and searching for the signal of new physics beyond the SM. The essential problem in the heavy hadron decays is to evaluate the matrix element of the effective Hamiltonian. 
Both in the semi-leptonic and nonleptonic decays, the heavy-to-light form factors play a crucial role in evaluating the decay amplitudes.  At leading power of the heavy quark expansion, the mesonic heavy-to-light form factors such as $B \to \pi$ form factor can be divided into two parts, the nonfactorizable soft form factor and the factorizable hard spectator scattering contribution. The latter is calculable but suppressed by the strong coupling constant. To study the soft form factor, the nonperturbative methods such as Lattice simulation and light-cone sum rules(LCSR) need to be employed. There exist two kinds of LCSRs, namely light-hadron LCSR \cite{Balitsky:1989ry,Belyaev:1993wp} and heavy-hadron LCSR \cite{Khodjamirian:2005ea,Khodjamirian:2006st,DeFazio:2005dx},  which are classed according to the hadron states (light hadron in  the final state or heavy hadron in the initial state, respectively) entering the correlation function.  In the LCSR framework, the soft form factors have been investigated using different correlation functions, and the QCD corrections as well as the resummation of large logarithmic terms have also been performed \cite{Khodjamirian:2011ub,
Wang:2015vgv, Shen:2016hyv, Wang:2017jow,Lu:2018cfc, Gao:2019lta}. For the baryonic form factors, the power counting is different from the mesonic case. It was found that at the leading power, the baryonic heavy-to-light form factor is completely factorizable \cite{Wang:2011uv}, but numerically  suppressed compared with the power suppressed soft contribution due to the $\alpha_s^2$  suppression  from hard gluon exchange. Therefore, one of the most important issues in the heavy baryon decays is to study the heavy-to-light form factors with non-perturbative method.

As more and more data of the heavy baryon decays are accumulated in the large hadron collider, it turns more important to precisely study the heavy baryon decays theoretically. The baryonic heavy-to-light form factors, such as $\Lambda_b \to \Lambda$ and $\Lambda_b \to p$,  have been studied using Lattice QCD simulaltion, the LCSRs and the light-front quark model, etc.\cite{Wang:2009hra, Khodjamirian:2011jp, Wang:2015ndk,Huang:2004vf,Zhao:2018zcb,Detmold:2016pkz,Aliev:2010uy,Feldmann:2011xf,Wei:2009np,Detmold:2015aaa}. The lattice simulation is based on the first principle of QCD and believed to be a reliable method, while it prefers to work in the small recoil region. The LCSR is applicable in the large recoil region, thus the LCSR predictions supplement to the Lattice result.  The light-baryon LCSR has been employed to study the baryonic heavy-to-light transition form factors in \cite{Khodjamirian:2011jp,Wang:2008sm,Huang:2004vf}, and the  $\Lambda_b \to p$ form factors were studied with $\Lambda_b$-LCSR with the Chernyak-Zhitnitsky(CZ) current being adopt as the interpolating current for the proton\cite{Wang:2009hra}, where the predictions of the $\Lambda_b \to p $ form factors are too small compared with the Lattice QCD and the LCSR with proton LCDAs  \cite{Khodjamirian:2011jp,Detmold:2015aaa}.

In this work, we will take advantage of the $\Lambda_b$-LCSR to  revisit the $\Lambda_b \to p $ form factors and simultaneously study the $\Lambda_b \to  N^\ast(1535)$(in what follows we will use $N^\ast$ to denote $N^\ast(1535)$ for convenience) transition form factors where $N^\ast$ $(1/2^-)$ is the parity conjugate state of the proton $(1/2^+)$.  Compared with the previous studies, we will make improvement in the following aspects:
\begin{itemize}
\item Since the interpolating current is not unique for the baryonic state, we will employ three different interpolating currents of the nucleon in the correlation functions, including the Ioffe current, the tensor current and the leading power(LP) current, so as to find out which one is the most appropriate choice in the $\Lambda_b$-LCSR by comparing our predictions with the experimental data or other theoretical results.
\item We will isolate  the proton state and $N^\ast$ state simultaneously in the hadronic representation of correlation functions. Thereby, there will be no redundant Lorentz structures in the partonic representation of the correlation function. We can then obtain all the $\Lambda_b \to p, N^\ast$ form factors without ambiguity by solving the equations of sum rules.
\item The light-cone distribution amplitudes (LCDAs) of $\Lambda_b$-baryon is the most important input in the sum rules of the form factors, which are not well determined yet. We will take advantage of the complete set of the three-particle LCDAs of $\Lambda_b$-baryon up to twist-5 in our calculation, some of which have been omitted in the previous studies. In addition, we will employ five different models for the LCDAs of $\Lambda_b$-baryon for a comparison \cite{Ball:2008fw,Ali:2012pn,Bell:2013tfa}.
\item We will obtain the leading power contribution of the universal $\Lambda_b \to p$ form factors by taking the SCET limit of the correlation function. By comparing the leading power contribution with the full result, we can estimate the size of the power suppressed contributions. 
\end{itemize}

Except for the theoretical improvements, the form factors of $\Lambda_b\to N^*$ have some phenomenological significance. CP violation has never been established in the baryon systems. Multi-body $\Lambda_b$ decays are among the most important processes to search for CP violation in baryons, for example $\Lambda_b\to p\pi^-\pi^+\pi^-$ \cite{LHCb:2016yco}. The intermediate states such as $N^*$ would contribute to these multibody decays, especially for the strong phases. In the theoretical calculations, the transition form factors of $\Lambda_b\to N^*$ are important input quantities. This case is similar to the theoretical prediction on the discovery channel of double-charm baryons, $\Xi_{cc}^{++}\to \Lambda_c^+K^-\pi^+\pi^+$, where the form factors of intermediate states are firstly calculated \cite{Yu:2017zst,Wang:2017mqp}.  

This paper is organized as follows. In the next section we start with the definition of the form factors and the symmetry relations in the heavy-quark/SCET limit,  then the sum rules for the  $\Lambda_b \to p, N^\ast(1535)$ form factors are obtained, and the investigation on the result from SCET limit is also given in this section. The details of the numerical analysis of the form factors are collected in section 3, and we also give our predictions on some observables for the the exclusive semileptonic decays $\Lambda_b \rightarrow p(N^*)l^-\bar{\nu_l}$. The last section is reserved for the summary. The paper also contains two appendices where the expressions of the LCSR for the form factors (App.A) and the models of LCDAs of $\Lambda_b$ baryon are collected (App.B).

\section {The LCSR of  $\Lambda_b \to p, N^\ast(1535)$ form factors }
\label{section: tree-level LCSR}

\subsection{ $\Lambda_b \to {p, N^\ast}$ form factors}
\label{subsection: form factor definition}
The  heavy-to-light $\Lambda_b \to  p$ and   $\Lambda_b \to  N^\ast$ form factors induced by $V-A$ current are defined as\cite{Mannel:1990vg}
\begin{eqnarray}
\left\langle p(p,s')\left|\bar{u} \gamma^{\mu} b\right| \Lambda_{b}(P,s)\right\rangle&=& \bar{u}\left(p, s'\right)\left(f_{1} \gamma^{\mu}+f_{2} i\sigma^{\mu\nu}\hat q_\nu +f_{3} \hat q^\mu\right) u\left(P, s\right), \nonumber \\ 
\left\langle p(p,s')\left|\bar{u} \gamma^{\mu} \gamma_{5} b\right| \Lambda_{b}(P,s)\right\rangle&=&  \bar{u}\left(p, s'\right)\left(g_{1} \gamma^{\mu}+g_{2} i \sigma^{\mu\nu}\hat q_\nu +g_{3} \hat q^\mu\right) \gamma_5u\left(P, s\right), \nonumber \\
\left\langle N^\ast(p,s')\left|\bar{u} \gamma^{\mu} b\right| \Lambda_{b}(P,s)\right\rangle&=& \bar{u}\left(p, s'\right)\left(F_{1} \gamma^{\mu}+F_{2} i\sigma^{\mu\nu}\hat q_\nu +F_{3} \hat q^\mu\right)\gamma_{5} u\left(P, s\right) ,  \nonumber \\
\left\langle N^\ast(p,s')\left|\bar{u} \gamma^{\mu} \gamma_{5} b\right| \Lambda_{b}(P,s)\right\rangle&=& \bar{u}\left(p, s'\right)\left(G_{1} \gamma^{\mu}+G_{2}i\sigma^{\mu\nu}\hat q_\nu +G_{3}\hat q^\mu\right)  u\left(P, s\right).
\end{eqnarray}
where $\sigma_{\mu\nu}=i[\gamma_{\mu},\gamma_{\nu}]/2$, and $u(P,s) \ (u(p,s^{'}))$ is the spinor of the $\Lambda_b$ baryon (proton or $N^*$) with the momentum $P\ (p)$ and the spin $s\ (s^{'})$. The form factors $f_i\ (F_i)$ and $g_i\ (G_i)$ depend on the invariant mass squared of the lepton pair $q^2$ with $ q_{\nu}=P_{\nu}-p_{\nu}$ , and $ \hat q_\nu=q_\nu/m_{\Lambda_b}$ so that all the form factors have the same mass dimension.
In HQET, one can use the heavy-baryon velocity $v^{\mu}=P^{\mu} / m_{\Lambda_{b}}$ to project the $b$ -quark field onto its large-spinor component $h_{v}^{(b)}=\not\! v h_{v}^{(b)}$, then the form factors can be written as
\begin{eqnarray}
\langle N^{(*)}(p,s')|{\bar{q}} \Gamma b | \Lambda_{b}(P,s)\rangle &\rightarrow& \langle N^{(*)}(p,s')|{\bar{q}} \Gamma h^{(b)}_v | \Lambda_{b}(v,s)\rangle \nonumber\\
&=&\bar{u}_{N^{(*)}}(p,s') \left[\zeta_{1}(q^{2})+\not\! v \zeta_{2}(q^{2})\right] \Gamma u_{\Lambda_{b}}(v,s),
\end{eqnarray}
where $N^{(*)}$ stands for proton or $N^{*}$, $\Gamma$ is any Dirac matrix, $|\Lambda_b(v,s)\rangle$ is the heavy-baryon state with the heavy-baryon spinor $u_{\Lambda_b}(v,s)=\not\! vu_{\Lambda_b}(v,s)$ in HQET. Therefor, $\zeta_{1}(q^{2})$ and $\zeta_{2}(q^{2})$ are the only two form factors that should be present at leading order in $\alpha_{s}$ and $\Lambda / m_{b}$ according to HQET.  Taking the heavy-quark limit (neglecting the contributions suppressed by $\Lambda / m_{b}$ and  $m_{N^{(\ast)}} / m_{\Lambda_b}$ ) in the definition of the form factors  yields:
\begin{eqnarray}
&&f_1=g_1=\zeta_1;   \,\,\,\,\, f_2=g_2=f_3=g_3=\zeta_2;   \nonumber \\
&&F_1=G_1=-\zeta_1;  \,\,\,\,\, F_2=G_2=F_3=G_3=\zeta_2. 
\label{hqlimit}
\end{eqnarray}
In the heavy-to-light transitions, the light quark moves approximately along the light-cone, thus it must be regarded as a collinear quark in the soft-collinear effective theory(SCET). In SCET\cite{Bauer:2000yr,Beneke:2002ph}, the collinear quark field is divided into the large component and small component, namely $q=\xi+\eta $ with $\xi=\frac{\not \bar n\not n }{ 4}q$ and two light-like unit vectors $\bar{n}^2=n^2=0$, satisfying $(n^{\mu}+\bar{n}^{\mu})=2v^{\mu}$ and $n\cdot \bar{n}=2$. At leading power, the large component $\xi$, $h_v^{(b)}$ will take the place of the $u$ quark field and $b$ quark field respectively, then the weak current turns to
\begin{eqnarray}
j_{\mu, V} =\bar \xi \, \not\! n \, h_v^{(b)}\frac{\bar n_{\mu}}{ 2},\ \ \  j_{\mu, A}=\bar \xi \, \not\! n \gamma_5 \, h_v^{(b)}\frac{\bar n_{\mu}}{2}\,.
\end{eqnarray}
The matrix element of the leading power current will result in one universal form factor due to the heavy quark symmetry and large recoil symmetry, i.e.,
\begin{eqnarray}
\langle N^{(*)}(p,s')|\bar \xi \Gamma h_v|\Lambda_{b}(v,s)\rangle& =&\bar{u}_{N^{(*)}}(p,s') \frac{\not n\not\bar n }{4} \left[\zeta_{1}(q^{2})+\not\! v \zeta_{2}(q^{2})\right]\Gamma u_{\Lambda_{b}}(v,s)\nonumber \\
&\to&   \xi(q^{2}) \bar{u}_{N^{(*)}}(p,s') \Gamma u_{\Lambda_{b}}(v,s).
\end{eqnarray}

\subsection{Interpolating currents and correlation function}
Following the standard strategy of heavy-hadron LCSR,  we start with the two-point correlation function sandwiched between vacuum and the on-shell $\Lambda_b$-baryon state:
\begin{eqnarray}
\Pi_{\mu, a}(p,q)= i \int d^4 x  \, e^{i p \cdot x} \, \langle 0 |T \{\eta(x), j_{\mu, a}(0) \}| \Lambda_b(p+q) \rangle \,,
\label{definition: correlator}
\end{eqnarray}
where the local current $\eta$ is the interpolating current for the light baryons $N^{(\ast)}$ and
the $j_{\mu, a}$ stands for the heavy-light weak transition current $\bar u \, \Gamma_{\mu, a} \, b $
with the index ``$a$" indicating a certain Lorentz structure, i.e.,
\begin{eqnarray}
j_{\mu, V} =\bar u \, \gamma_{\mu} \, b\,, &  \qquad & j_{\mu, A}=\bar u \, \gamma_{\mu}\, \gamma_5 \, b\,.
\end{eqnarray}

For comparison, we adopt three types of the interpolating current for  the $N^{(\ast)}$ baryon, namely the Ioffe current, the tensor current and the LP current respectively\cite{,Ioffe:1982ce,Braun:2006hz}, and they are listed below
\begin{eqnarray}\eta_{\rm IO}(x)&=&\varepsilon^{a b c}\left[u^{a T}(x) C \gamma^{\rho} u^{b}(x)\right] \gamma_5 \gamma_{\rho} d^{c}(x)\nonumber, \\
\eta_{\rm TE}(x)&=&\varepsilon^{a b c}\left[u^{a T}(x) C \sigma^{\rho\sigma} u^{b}(x)\right] \gamma_5 \sigma_{\rho\sigma} d^{c}(x)\nonumber, \\
\eta_{\rm LP}(x)&=&\varepsilon^{a b c}\left[u^{a T}(x) C \not\!n u^{b}(x)\right] \gamma_5 \not\!n d^{c}(x).
\end{eqnarray}
where $T$ represents the transposition, $C$ is the charge conjugation matrix and the sum runs over the color indices $a,b,c$. 

Following  the usual procedure of LCSR, the correlation function is to be expressed in the hadronic representation as well as calculated by the operator product expansion in QCD. On the hadronic level, one inserts a complete set of states with the same quantum numbers of light baryon $N^{(\ast)}$ between the interpolating current $\eta_i$ and the transition current $j_{\mu, a}$. In order to obtain the $\Lambda_b \rightarrow p, N^{*}$ form factors without ambiguity, we isolate the contribution both from the lowest state $p$ and its negative-parity partner $N^{\ast}$ in the hadrnoic dispersion relation. The results in two different representations of the correlation function are then matched by employing the quark-hadron duality assumption. Finally, one needs to perform the Borel-transformation to eliminate the possible subtraction terms which may appear in the dispersion relation  and to suppress the contributions from the higher states.

At hadronic level, the correlation function can be expressed in terms of hadronic matrix elements as follows
\begin{eqnarray}
\Pi_{\mu,a}(p,q)=&& \frac{1}{m_{p}^2-p^{ 2}}\sum_{s'}\left\langle 0\left|\eta_i(0)\right| p\left(p,s'\right)\right\rangle\left\langle p\left(p,s'\right)\left| j_{\mu, a}(0)\right| \Lambda_b(p+q)\right\rangle \nonumber \\
&&+\frac{1}{m_{N^{\ast}}^2-p^{ 2}}\sum_{s'}\left\langle 0\left|\eta_i(0)\right| N^{\ast}\left(p,s'\right)\right\rangle\left\langle N^{\ast}\left(p,s'\right)\left| j_{\mu, a}(0)\right| \Lambda_b(p+q)\right\rangle+...\  ,
\end{eqnarray}
where the ellipsis stands for the contribution of excited and continuum states with the same quantum numbers as $p$ and $N^*$. The couplings of
the $p, N^\ast$ with the interpolating currents $\eta$ (the decay constants) are defined as
\begin{eqnarray}
&&\langle 0| \eta_{\rm IO} | p(p,s')\rangle = \lambda_{1}^p m_p u\left(p,s'\right),\,\,\,\,\,\ \ \ \ 
\langle 0| \eta_{\rm IO} | N^\ast(p,s')\rangle = \lambda^{N^\ast}_{1} m_{N^\ast} \gamma_5u\left(p,s'\right),
\nonumber \\
&&\langle 0| \eta_{\rm TE} | p(p,s')\rangle = \lambda^{p}_{2} m_{p} u\left(p,s' \right),\,
\,\,\,\ \ \ \ 
\langle 0| \eta_{\rm TE} | N^\ast(p,s')\rangle = \lambda^{N^\ast}_{2} m_{N^\ast} \gamma_5u\left(p,s'\right),
\nonumber \\
&&\langle 0| \eta_{\rm LP} | p(p,s')\rangle = \lambda^{p}_{3} \not\! n n\cdot p u\left(p,s'\right),\,
\,\,\,
\langle 0| \eta_{\rm LP} | N^\ast(p,s')\rangle = \lambda^{N^\ast}_{3} \not\! n n\cdot p \gamma_5u\left(p,s'\right).
\end{eqnarray}

By using the equation of motion,  the correlation function can be decomposed into 12 independent, Lorentz invariant amplitudes. For the vector current $j_{\mu , V}$, we have:
 \begin{eqnarray}
 	\Pi^{i}_{\mu,V}(p,q)=\left(\Pi^i_1p_{\mu}+\Pi^i_2p_{\mu}\not\!q+\Pi^i_3\gamma_{\mu}
 +\Pi^i_4\gamma_{\mu}\not\!q+\Pi^i_5q_{\mu}+\Pi^i_6q_{\mu}\not\!q\right)u_{\Lambda_b}(p+q),
\end{eqnarray}\label{eq:Pii}
where the dependence of $\Pi^i_{1-6}$ on $p^2 , q^2$ is not shown specifically. A similar decomposition can be done for the axial-vector current $j_{\mu ,A}$:
 \begin{eqnarray}
 	\Pi^{i}_{\mu,A}(p,q)=\left(\tilde\Pi^i_1p_{\mu}+\tilde\Pi^i_2p_{\mu}\not\!q
 +\tilde\Pi^i_3\gamma_{\mu}+\tilde\Pi^i_4\gamma_{\mu}\not\!q+\tilde\Pi^i_5q_{\mu}+\tilde\Pi^i_6q_{\mu}\not\!q\right)\gamma_5 u_{\Lambda_b}(p+q).
 \end{eqnarray}
Employing the definitions of the decay constants and the form factors, and summing over the spin of the $p$ and $N^*$, we arrive at the hadronic  expression for the correlation functions. For the vector current $j_{\mu , V}$, we have:
\begin{eqnarray}
	\Pi^i_{\mu , V}&=&\frac{\lambda^p_im_p}{m^2_p-p^2}\Big\{2f_1(q^2)p_{\mu}-\frac{2 f_2(q^2)}{m_{\Lambda_b}}p_{\mu}\not\!q \nonumber  \\
	&+&(m_{\Lambda_b}-m_p)\Big[\frac{m_{\Lambda_b}+m_p}{m_{\Lambda_b}}f_2(q^2)-f_1(q^2)\Big]\gamma_{\mu}
+\Big[f_1(q^2)-\frac{m_{\Lambda_b}+m_p}{m_{\Lambda_b}}f_2(q^2)\Big]\gamma_{\mu}\not\!q \nonumber  \\
	&+&\frac{m_{\Lambda_b}+m_p}{m_{\Lambda_b}}[f_2(q^2)+f_3(q^2)]q_{\mu}
-\frac{f_2(q^2)+f_3(q^2)}{m_{\Lambda_b}}q_{\mu}\not\!q\Big\}u_{\Lambda_b}(p+q) \nonumber \\
	&+&\frac{\lambda^{N^*}_im_{N^*}}{m^2_{N^*}-p^2}\Big\{2F_1(q^2)p_{\mu}
+\frac{2 F_2(q^2)}{m_{\Lambda_b}}p_{\mu}\not\!q \nonumber  \\
	&-&(m_{\Lambda_b}+m_{N^*})\Big[\frac{m_{\Lambda_b}-m_{N^*}}{m_{\Lambda_b}}F_2(q^2)+F_1(q^2)\Big]\gamma_{\mu}
+\Big[F_1(q^2)-\frac{m_{N^*}-m_{\Lambda_b}}{m_{\Lambda_b}}F_2(q^2)\Big]\gamma_{\mu}\not\!q \nonumber  \\
	&-&\frac{m_{\Lambda_b}-m_{N^*}}{m_{\Lambda_b}}[F_2(q^2)+F_3(q^2)]q_{\mu}
+\frac{F_2(q^2)+F_3(q^2)}{m_{\Lambda_b}}q_{\mu}\not\!q\Big\}u_{\Lambda_b}(p+q) \nonumber \\
	&+&\int_{s_0^h}^{\infty} \frac{ds}{s-p^2}\Big(\rho_1^i(s,q^2)p_{\mu}+\rho_2^i(s,q^2)p_{\mu}\not\!q \nonumber \\ &+&\rho_3^i(s,q^2)\gamma_{\mu}+\rho_4^i(s,q^2)\gamma_{\mu}\not\!q+\rho_5^i(s,q^2)q_{\mu}
+\rho_6^i(s,q^2)q_{\mu}\not\!q\Big).
\label{hadron}
\end{eqnarray}
where $i$ denotes the interpolating current, the contribution of the excited and continuum states is described by the spectral densities $\rho_j^i$ with $j=1,...,6$ , and $s_0^h$ is the corresponding threshold parameter. From the above expression, the invariant amplitudes $\Pi^i_{1,..6}$  can be directly related to the form factors which are then extracted after the $\Pi^i_{1,..6}$ being calculated in QCD . The similar expression for the invariant amplitude $\tilde\Pi^{i}_{1,..,6} $ of the axial current $j_{\mu, A}$ can be obtained from Eq.(\ref{hadron}) by replacing $f_i \rightarrow g_i , F_i\rightarrow G_i$, changing the sign of $m_{\Lambda_b}$ and adding the $\gamma_5$ before the spinor $u_{\Lambda_b}(p+q)$.

\subsection{The light-cone sum rules for the form factors}

Now we turn to compute the correlation function $\Pi_{i\mu, a}(p, q)$
form the partonic side. For the space-like interpolating momentum $|\bar n \cdot p| \sim {\cal O} (\Lambda)$
and $n \cdot p\sim m_{\Lambda_b}$ , the product of the two currents can be expanded around the light-cone $x^2\simeq 0$. After contracting the $u$ quark field and integrating out the coordinate, the correlation function at partonic level  can be factorized into the convolution of the hard kernel and the LCDA of $\Lambda_b$ baryon, as shown in the diagram of Fig.\ref{fig:tree_correlator} :

\begin{figure}
	\begin{center}
		\includegraphics[width=0.4 \columnwidth]{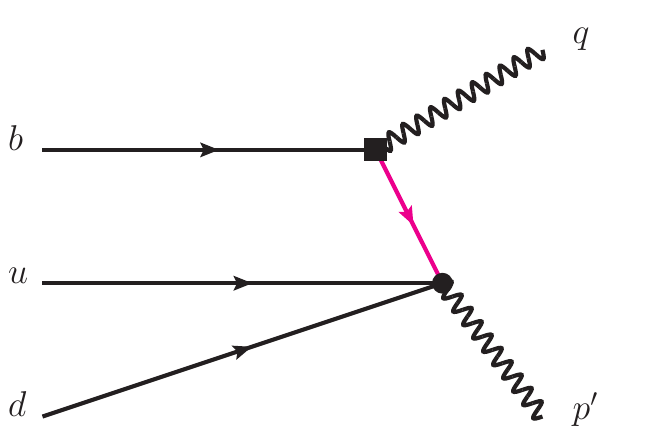}
		\vspace*{0.1cm}
		\caption{Diagrammatical representation of the correlation function
			$\Pi_{\mu, a}(n \cdot p,\bar n \cdot p)$ at tree level,
			where the black square denotes the weak transition vertex, the black blob represents
			the Dirac structure of the $N^{(\ast)}$-baryon current and the pink internal line indicates the hard-collinear
			propagator of the $u$ quark. }
		\label{fig:tree_correlator}
	\end{center}
\end{figure}
\begin{eqnarray}
\Pi^{\rm }_{i\mu, a}(p, q)=\int d \omega_1^{\prime} \int d \omega_2^{\prime} \,
T^{(0),a}_{\alpha \beta \gamma \delta}(p, q, \omega_1^{\prime}, \omega_2^{\prime})\,
\Phi_{\Lambda_b}^{ \, \alpha \beta \delta}(\omega_1^{\prime}, \omega_2^{\prime}) \,,
\label{tree-level factorization at partonic level}
\end{eqnarray}
where the superscript (0) indicates the tree-level approximation, $\omega'_{i}(i=1.2)$ are the energies of the $u-$ and $d-$quarks, the Lorenz index $\mu$ is not shown exactly, and $\alpha ,\beta , \gamma,\delta$ are the Dirac indices.

Evaluating the diagram in Fig. \ref{fig:tree_correlator} leads to the leading-order hard kernel
\begin{eqnarray}
	T^{(\rm IO,0)}_{\alpha \beta \gamma \delta}(p,q)&=&-\left[2C\gamma^\rho{\not\! k\over k^2}\gamma_\mu(I,\gamma_5)\right]_{\alpha\delta} (\gamma_5\gamma_\rho)_{\gamma\beta}, \nonumber \\
	T^{(\rm Te,0)}_{\alpha \beta \gamma \delta}(p,q)&=&-\left[2C\sigma^{\rho\sigma}{\not\! k\over k^2}\gamma_\mu(I,\gamma_5)\right]_{\alpha\delta} (\gamma_5\sigma_{\rho\sigma})_{\gamma\beta},\nonumber \\
	T^{(\rm LP,0)}_{\alpha \beta \gamma \delta}(p,q)&=&-\left[2C\not\! n{\not\! k\over k^2}\gamma_\mu(I,\gamma_5)\right]_{\alpha\delta} (\gamma_5\not\! n)_{\gamma\beta} ,
	\label{tree-level factorization at partonic level}
\end{eqnarray} and the definition of the general light-cone hadronic matrix element
in coordinate space \cite{Bell:2013tfa} is given by
\begin{eqnarray}
	\Phi_{\Lambda_b}^{\alpha \beta \delta}(t_1, t_2)
	&\equiv&  \epsilon_{i j k} \, \langle 0 | \left [u^{\rm T}_{i} (t_1 \bar n) \right ]_{\alpha} \,
	[0, t_1 \bar n] \, \left [d_{j} (t_2 \bar n) \right ]_{\beta} \, [0, t_2 \bar n] \,
	\left [ b_{k}(0)\right ]_{\delta} | \Lambda_b(v) \rangle \nonumber \\
	&=&  \frac{1}{4 } \, \left \{ f_{\Lambda_b}^{(1)}(\mu) \,
	\left [ \tilde{M}_1(v, t_1, t_2) \, \gamma_5 \, C^{T} \right ]_{\beta \alpha}
	+  f_{\Lambda_b}^{(2)}(\mu) \,
	\left [ \tilde{M}_2(v, t_1, t_2) \, \gamma_5 \, C^{T} \right ]_{\beta \alpha} \right \} \,
	\left [ \Lambda_b(v) \right ]_{\delta}   \,, \nonumber \\
\end{eqnarray}
where  the $b$-quark field needs to be understood as an effective heavy field in HQET , $t_i$ are the real numbers which describe the location of the valance quarks inside the $\Lambda_b$ baryon on the light-cone. Performing the Fourier transformation, the momentum space light-cone projectors of the LCDAs of $\Lambda_b$ baryon in $D$ dimensions read
\begin{eqnarray}
M_2(\omega_1^{\prime},\omega_2^{\prime}) &=& \frac {\! \not n}{2} \, \psi_2(\omega_1^{\prime},\omega_2^{\prime})
+ \frac {\! \not \bar  n}{2} \, \psi_4(\omega_1^{\prime},\omega_2^{\prime}) \nonumber
\\
&& -\frac{1}{D-2} \, \gamma_{\perp}^{\mu} \, \left  [ \psi_{\perp, 1}^{+-}(\omega_1^{\prime},\omega_2^{\prime}) \,
\frac{\! \not n  \, \! \not \bar  n}{4}  \, \frac{\partial}{\partial k_{1 \perp}^{\mu}}
+ \psi_{\perp, 1}^{-+}(\omega_1^{\prime},\omega_2^{\prime}) \,
\frac{\! \not \bar n  \, \! \not  n}{4}  \, \frac{\partial}{\partial k_{1 \perp}^{\mu}} \right  ] \nonumber
\\
&& -\frac{1}{D-2} \, \gamma_{\perp}^{\mu} \, \left  [ \psi_{\perp, 2}^{+-}(\omega_2^{\prime},\omega_2^{\prime}) \,
\frac{\! \not n  \, \! \not \bar  n}{4}  \, \frac{\partial}{\partial k_{2 \perp}^{\mu}}
+ \psi_{\perp, 2}^{-+}(\omega_1^{\prime},\omega_2^{\prime}) \,
\frac{\! \not \bar n  \, \! \not  n}{4}  \, \frac{\partial}{\partial k_{2 \perp}^{\mu}} \right  ]  \,,
\label{chiral-odd projector}\\
M_1(\omega_1^{\prime},\omega_2^{\prime}) &=& \frac{\! \not \bar  n \, \! \not  n }{8} \,
\psi_{3}^{+-}(\omega_1^{\prime},\omega_2^{\prime})
+ \frac{\! \not  n  \, \! \not \bar  n}{8} \,
\psi_{3}^{-+}(\omega_1^{\prime},\omega_2^{\prime}) \nonumber
\\
&& - \frac{1}{D-2} \left [ \psi_{\perp, 3}^{(1)}(\omega_1^{\prime},\omega_2^{\prime}) \! \not  v
\, \gamma_{\perp}^{\mu}  \, \frac{\partial}{\partial k_{1 \perp}^{\mu}}
+ \psi_{\perp, 3}^{(2)}(\omega_1^{\prime},\omega_2^{\prime}) \, \gamma_{\perp}^{\mu}  \,
\! \not  v \, \frac{\partial}{\partial k_{2 \perp}^{\mu}}  \right ] \nonumber
\\
&& - \frac{1}{D-2} \left [ \psi_{\perp, Y}^{(1)}(\omega_1^{\prime},\omega_2^{\prime}) \! \not \bar n
\, \gamma_{\perp}^{\mu}  \, \frac{\partial}{\partial k_{1 \perp}^{\mu}}
+ \psi_{\perp, Y}^{(2)}(\omega_1^{\prime},\omega_2^{\prime}) \,  \gamma_{\perp}^{\mu}  \,
\! \not  \bar n \, \frac{\partial}{\partial k_{2 \perp}^{\mu}}  \right ] \,,
\label{chiral-even projector}
\end{eqnarray}
where we have adjusted the notation of the $\Lambda_b$-baryon LCDA defined in \cite{Bell:2013tfa}. Applying the equations of motion in the Wandzura-Wilczek approximation yields
\begin{eqnarray}
\psi_{\perp, 1}^{-+}(\omega_1^{\prime},\omega_2^{\prime})=\omega_1^{\prime} \,
\psi_4(\omega_1^{\prime},\omega_2^{\prime}) \,, \qquad
\psi_{\perp, 2}^{+-}(\omega_1^{\prime},\omega_2^{\prime})=\omega_2^{\prime} \,
\psi_4(\omega_1^{\prime},\omega_2^{\prime})  \,.
\label{EOM for Lambdab DA}
\end{eqnarray}
For convenience, we introduce two new parameters $\omega , u$ which satisfy  $\omega=\omega_1'+\omega_2', u=\omega_1'/\omega$. Combining the projectors of the LCDAs of $\Lambda_b$ baryon and the hard kernels and simplifying the Lorentz structures, the invariant amplitudes at the partonic side can be obtained as 
\begin{eqnarray}\label{EQ:invaramp}
	\Pi^i_j(p^2 , q^2)=\sum_{n=1,2}\int^{\infty}_0d\omega\int^1_0du\frac{w^{(i)}_{jn}(p^2, q^2)}{D^n},
\end{eqnarray}
with the denominator
\begin{eqnarray}
D=\bar{\sigma}p^2+\sigma q^2-\sigma\bar{\sigma}m^2_{\Lambda_b},
\end{eqnarray}
where  $\sigma={\omega}/{m_{\Lambda_b}} ,\  \bar{\sigma}=1-\sigma$. The functions $w^{(i)}_{jn}$ are distinguished by their indices: i=IO, TE, LP(the interpolating current), $j=1,..,6$(the invariant amplitude) and $n=1,2$(the power of the denominator). 

 The invariant amplitudes in Eq.(\ref{EQ:invaramp}) can be expressed by the following dispersion integral
\begin{eqnarray}
\Pi^{i}_{j}(p^2 , q^2)=\frac{1}{\pi} \int^{\infty}_0 \frac{ds}{s-p^2} {\rm Im}_s \Pi^i_j(s,q^2),
\end{eqnarray}
 Compared with the hadronic dispersion integral involved the $p,N^\ast$ and higher states in Eq.(\ref{hadron}), it can be employed to extract the form factors. Using the quark-hadron duality to approximate the contribution of the hadronic states above the threshold $s^h_0$ :
\begin{eqnarray}
	\int^{\infty}_{s^h_0}\frac{ds}{s-p^2}\rho^i_j(s,q^2)\approx\frac{1}{\pi}\int^{\infty}_{s_0}\frac{ds}{s-p^2}{\rm Im}_s \Pi^{i}_j(s,q^2),
\end{eqnarray}
where $s_0$ is the effective threshold parameter. After performing the Borel transformation, we can get the sum rules for the form factors of the vector transition current $j_{\mu, V}$:
\begin{eqnarray}
	f_1(q^2)&=&\frac{e^{m_p^2/M^2}}{2m_p(m_p+m_{N^*})\lambda_i^p}\frac{1}{\pi}\int^{s_0}_0 ds e^{-s/M^2}\Big[(m_{\Lambda_b}+m_p)\Big({\rm Im}_s\Pi^i_1(s,q^2) \nonumber \\
	&+&(m_{\Lambda_b}-m_{N^*}){\rm Im}_s\Pi^i_2(s,q^2)\Big)+2{\rm Im}_s\Pi^i_3(s,q^2)+2(m_{N^*}-m_p){\rm Im}_s\Pi^i_4(s,q^2)\Big],	
 \nonumber \\
	f_2(q^2)&=&\frac{m_{\Lambda_b}e^{m_p^2/M^2}}{2m_p(m_p+m_{N^*})\lambda_i^p}\frac{1}{\pi}\int^{s_0}_0 ds e^{-s/M^2}\Big[{\rm Im}_s\Pi^i_1(s,q^2)
	+(m_{\Lambda_b}-m_{N^*}){\rm Im}_s\Pi^i_2(s,q^2)\nonumber \\&-&2{\rm Im}_s\Pi^i_4(s,q^2)\Big],	
 \nonumber \\
	f_3(q^2)&=&-\frac{e^{m_p^2/M^2}}{2m_p(m_p+m_{N^*})\lambda_i^p}\frac{1}{\pi}\int^{s_0}_0 ds e^{-s/M^2}\Big[{\rm Im}_s\Pi^i_1(s,q^2)+(m_{\Lambda_b}-m_{N^*}){\rm Im}_s\Pi^i_2(s,q^2) \nonumber \\
	&-&2{\rm Im}_s\Pi^i_4(s,q^2)-2{\rm Im}_s\Pi^i_5(s,q^2)-2(m_{\Lambda_b}-m_{N^*}){\rm Im}_s\Pi^i_6(s,q^2)\Big],
	\label{fi}
\end{eqnarray}

\begin{eqnarray}
	F_1(q^2)&=&-\frac{e^{m_{N^*}^2/M^2}}{2m_{N^*}(m_p+m_{N^*})\lambda_i^{N^*}}\frac{1}{\pi}\int^{s_0}_0 ds e^{-s/M^2}\Big[(m_{\Lambda_b}-m_{N^*})\Big({\rm Im}_s\Pi^i_1(s,q^2) \nonumber \\
	&+&(m_{\Lambda_b}+m_{p}){\rm Im}_s\Pi^i_2(s,q^2)\Big)+2{\rm Im}_s\Pi^i_3(s,q^2)+2(m_{N^*}-m_p){\rm Im}_s\Pi^i_4(s,q^2)\Big]	,
 \nonumber \\ 
 F_2(q^2)&=&\frac{m_{\Lambda_b}e^{m_{N^*}^2/M^2}}{2m_{N^*}(m_p+m_{N^*})\lambda_i^{N^*}}\frac{1}{\pi}\int^{s_0}_0 ds e^{-s/M^2}\Big[{\rm Im}_s\Pi^i_1(s,q^2)
	+(m_{\Lambda_b}+m_{p})Im_s\Pi^i_2(s,q^2)\nonumber \\&-&2{\rm Im}_s\Pi^i_4(s,q^2)\Big] ,	
 \nonumber \\
	F_3(q^2)&=&-\frac{e^{m_{N^*}^2/M^2}}{2m_{N^*}(m_p+m_{N^*})\lambda_i^{N^*}}\frac{1}{\pi}\int^{s_0}_0 ds e^{-s/M^2}\Big[{\rm Im}_s\Pi^i_1(s,q^2)+(m_{\Lambda_b}+m_{p}){\rm Im}_s\Pi^i_2(s,q^2) \nonumber \\
	&-&2{\rm Im}_s\Pi^i_4(s,q^2)-2Im_s\Pi^i_5(s,q^2)-2(m_{\Lambda_b}+m_{N^*}){\rm Im}_s\Pi^i_6(s,q^2)\Big] ,	
	\label{Fi}
\end{eqnarray}
The sum rules for the form factors $g_i(q^2)$ and $G_i(q^2)$ of the axial-vector transition current $j_{\mu ,A}$ can be obtained from the above expressions for the form factors $f_i(q^2)$ and $F_i(q^2)$ by replacing ${\rm Im}_s \Pi^i_j(s,q^2)\rightarrow {\rm Im}_s\tilde\Pi^i_j(s,q^2)$ respectively and changing the sign of $m_{\Lambda_b}$. In practice, to obtain the final result, we should integrate out the variable $s$.  All of the three procedures, including the expression of the correlation function with the dispersion integral, subtraction of continuum states with quark-hadron duality and Borel transformation, can be realized with the following substitution rules
\begin{eqnarray}
	\int d\omega\int du\frac{w(s , q^2)}{D}&\rightarrow& -m_{\Lambda_b} \int^1_0 du\int^{\sigma_0}_{0}\frac{d\sigma}{\bar{\sigma}}\omega(s,q^2)\exp\Big({-\frac{s}{M^2}}\Big) , \nonumber \\
	\int d\omega\int du\frac{w(s , q^2)}{D^2}&\rightarrow& \frac{m_{\Lambda_b}}{M^2} \int^1_0 du\int^{\sigma_0}_{0}\frac{d\sigma}{\bar{\sigma}^2}\omega(s,q^2)\exp\Big({-\frac{s}{M^2}}\Big)+\int^1_0 du\frac{\omega(s_0,q^2)\eta(\sigma_0)e^{-s_0/M^2}}{\bar{\sigma}^2_0m_{\Lambda_b}} , \nonumber \\
\end{eqnarray}
  where the transformed coefficient functions $\omega^{(i)}_{jn}(s,q^2)$ are presented in Appendix \ref{sec:AppendixA} and the involved parameters are defined as:
	\begin{eqnarray}
s(\sigma)=\sigma m_{\Lambda_b}^2-{\sigma q^2\over\bar \sigma},\,\,\, \eta(\sigma)=\left(1-{ q^2\over \bar \sigma^2m_{\Lambda_b}^2}\right)^{-1},
\end{eqnarray}
 and $\sigma_0$ is the positive solution of the corresponding quadratic equation for $s=s_0$:
 \begin{eqnarray}
     \sigma_0=\frac{(s_0+m^2_{\Lambda_b}-q^2)-\sqrt{(s_0+m^2_{\Lambda_b}-q^2)^2-4m^2_{\Lambda_b}s_0}}{2m^2_{\Lambda_b}}.
 \end{eqnarray}

\subsection{The SCET limit}

In this section, we present a short discussion on the SCET limit of the $\Lambda_b \to p$ form factors. We have mentioned that in the SCET limit, the three $\Lambda_b \to p$ form factors  reduce to a unique one. In order to obtain the unique form factor, the interpolating current should be fixed  so that the power suppressed part is eliminated and it is easy to find that the LP current will give rise to the leading power contribution. 
At leading power, the quark fields in the LP current should be replaced by the large component of the collinear quark fields
\begin{equation}\eta^{\rm LP}(x)=\varepsilon^{a b c}\left[\xi_u^{a T}(x) C \not\! n \xi_u^{b}(x)\right] \gamma_5\not \! n \xi_d^{c}(x) ,
\end{equation}
In addition, the weak current $\bar u \gamma_\mu(\gamma_5) b$ also  should be changed into $\bar \xi \not\! n(\gamma_5) h_v^{(b)}$, then the leading power hard kernel for the vector current can be obtained as
\begin{eqnarray}
T^{(0)}_{\alpha \beta \gamma \delta}(p,q)=-{1\over \bar n\cdot(p-k)}\left[C\not\! n{\not\!\bar n\over 2}\not\! n\right]_{\alpha\delta} (\gamma_5\not\! n)_{\gamma\beta} \,,
\label{tree-level factorization at partonic level}
\end{eqnarray}
Note that an additional $\gamma_5$ will appear for the axial vector current. The correlation function is then expressed  in terms of the convolution of the hard function and  the LCDAs of $\Lambda_b$ baryon
\begin{eqnarray}
\Pi^{\rm par,LP}_{\mu, a}(p, q)=f_{\Lambda_b}^{(2)}\int_0^\infty \omega d\omega {\bar\psi_4(\omega)\over \bar n\cdot p-\omega}\not\! n\Lambda_b(v,s), \end{eqnarray}
where $\bar\psi_4(\omega)=\omega \int_0^1 du\ \psi_4(u\omega,\bar u\omega)$. On the hardonic side, we can insert a complete set of the baryonic states with the same quark structure with the interpolating current to the correlation function and isolate the proton state(note that we do not isolate the negative parity state here, since there is no redundant Lorentz structures on the partonic side).  Combined with the hadronic representation of the correlation function, the LCSR for the leading power current  can be obtained after taking advantage of the quark-hardon duality and performing the Borel transform,  and the result reads
\begin{eqnarray}
\xi(n\cdot p)&=&{f_{\Lambda_b}^{(2)}\over  \lambda_p^{(3)}n\cdot p}{\rm Exp}\left( m_p^2 \over n\cdot p \omega_M\right)\int_0^{\omega_s} d\omega  e^{-\omega/ \omega_M}\bar\psi_4(\omega) .
\end{eqnarray}
where $n\cdot p\simeq (m^2_{\Lambda_b}+m_p^2-q^2)/m_{\Lambda_b}$ at large hadronic recoil, $M^2\equiv n\cdot p\omega_M$ and $s_0\equiv n\cdot p\omega_s$. This result is quite similar to the $\Lambda_b\to \Lambda$ form factor obtained in \cite{Wang:2015ndk}, and only the LCDA $\psi_4(u,\omega)$ appears in the LCSRs, which makes it straightforward to perform the next-to-leading order corrections to the sum rules at leading power, which will be left for a future study. 

\section{Numerical analysis}
\subsection{The results for form factors}\label{sec:FF}

 In this subsection, we present the numerical results of the form factors of  $\Lambda_b\rightarrow p(N^*)$ transition with the LCDA models of $\Lambda_b$-baryon given in Appendix B. The masses of the involved baryons are taken from\cite{PDG}: $m_{\Lambda_b}=5.620$ GeV, $m_p=0.938$ Gev, and $m_{N^*}=1.530$ GeV. The normalization parameters $\lambda^{p(N^*)}_i$ of the $p, N^*$ with the interpolating currents $\eta_i$  have been calculated from lattice-QCD\cite{Braun:2014wpa} , and the couplings $f^{(1,2)}_{\Lambda_b}$ are taken from the NLO QCD sum rule analysis\cite{Groote:1997yr}:
\begin{eqnarray}
	&\lambda^p_1 =-(3.877\pm 0.189)\times 10^{-2}\ \text{GeV}^{2}, \,\,\,\,
	\lambda^{N^*}_1=(2.463\pm 0.110)\times 10^{-2}\ \text{GeV}^{2}, \nonumber\\
	&\lambda^p_2=(7.764\pm 0.375)\times 10^{-2}\ \text{GeV}^{2},   \,\,\,\,\,\ \
    \lambda^{N^*}_2=(5.552\pm 0.278)\times 10^{-2}\ \text{GeV}^{2},\nonumber\\
	&\lambda^p_3 =(3.071\pm 0.360)\times 10^{-3}\ \text{GeV}^2,\,\,\,\,\ \
    \lambda^{N^*}_3=(0.757\pm 0.040)\times 10^{-3}\ \text{GeV}^2 , \nonumber\\
    &f^{(2)}_{\Lambda_b}=f^{(1)}_{\Lambda_b}=0.030\pm 0.005\   \text{GeV}^3 ,
\end{eqnarray}
 The sum rules for the form factors contain  two additional auxiliary parameters: the Borel parameter $M^2$ and the effective threshold parameter $s_0$, and the results of the form factors should be actually independent on these parameters. Therefore, one should find appropriate regions where the form factors  weakly depend on these parameters. In addition, the parameters $s_0$ and $M^2$ are adjusted in order to suppress the contributions of the continuum states and the high twist $\Lambda_b$-LCDA. To fulfill these requirements, the allowed regions  of the Borel parameter $M^2$ and the effective threshold parameters $s_0$ are chosen as 
\begin{eqnarray}
	&M^2=(2.6\pm 0.4)\ \text{GeV}^{2},\ \ \ \ \ 
	s_0=(2.56\pm 0.10)\ \text{GeV}^{2} ,
\end{eqnarray}
where the obtained regions of $M^2$ and $s_0$ are approximately in agreement with that used in \cite{Wang:2009hra}. 

Inputting the values of the parameters given above into the analytic expression of the LCSRs of the $\Lambda_b \rightarrow p$ and $\Lambda_b \rightarrow N^*$  form factors,  the numerical results of these form factors at $q^2=0$ from LCSRs are collected in  Table.\ref{form factors at q0:proton} and Table.\ref{form factors at q0:Nstart} respectively,  where three kinds of the interpolating current namely, Ioffe current, Tensor current and LP current are adopted in the correlation functions for a comparison,  and five different LCDA models of $\Lambda_b$-baryon (Gegenbauer-1, Gegenbauer-2, QCDSR, Exponential and Free-parton) are employed in the calculation of the correlation function in the partonic level in order that one has the chance to find out which one is the most preferable by comparing with the results of the experiment or the other studies. The uncertainties of the from factors arise from the variation of the thresholds parameter $s_0$, the Borel parameter $M^2$, the normalization constants for $p(N^*)$ and $\Lambda_b$-baryon and  the parameters in the LCDA models of $\Lambda_b$-baryon. 
\begin{table}
\centering
 \caption{The results of form factors $f_i$ and $g_i$ for $\Lambda_b \rightarrow p$ at $q^2=0$ under three interpolating currents for the proton(Ioffe, Tensor and LP) with five different LCDA models of $\Lambda_{b}$ baryon(Gegenbauer-1, Gegenbauer-2, QCDSR, Exponential and Free-parton) in this work. The results for the form factor $f_2(0)$ and $f_3(0)$ are same, $f_2(0)=f_3(0)$, as well as for $g_2(0)$ and $g_3(0)$. For comparison, we also collect the results of the form factors from other works.}
 \begin{tabular}{c c c c c }
 	\hline
 	\hline
 	&$f_1$&$f_2$&$g_1$&$g_2$\\
 	\hline
 	\textbf{Ioffe current} \\
 	Gegenbauer-1&$0.53\pm0.39$&$-0.12\pm0.099$&$0.53\pm0.39$&$-0.12\pm0.099$\\
 	Gegenbauer-2&$0.50\pm0.077$&$-0.11\pm0.021$&$0.50\pm0.077$&$-0.11\pm0.021$\\
 	QCDSR&$0.13\pm0.023$&$-0.023\pm0.004$&$0.13\pm0.023$&$-0.023\pm0.004$\\
 	Exponential&$0.14\pm0.087$&$-0.026\pm0.017$&$0.14\pm0.087$&$-0.026\pm0.017$\\
 	Free-parton&$0.17\pm0.11$&$-0.031\pm0.022$&$0.17\pm0.11$&$-0.031\pm0.022$\\
 	\hline
 	\textbf{Tensor current} \\
 	Gegenbauer-1&$0.37\pm0.34$&$-0.070\pm0.058$&$0.37\pm0.34$&$-0.070\pm0.058$\\
 	Gegenbauer-2&$0.36\pm0.079$&$-0.068\pm0.015$&$0.36\pm0.079$&$-0.068\pm0.015$\\
 	QCDSR&$0.11\pm0.023$&$-0.023\pm0.005$&$0.11\pm0.023$&$-0.023\pm0.005$\\
 	Exponential&$0.12\pm0.071$&$-0.024\pm0.014$&$0.12\pm0.071$&$-0.024\pm0.014$\\
 	Free-parton&$0.16\pm0.10$&$-0.033\pm0.021$&$0.16\pm0.10$&$-0.033\pm0.021$\\
 	\hline
 	\textbf{LP current} \\
 	Gegenbauer-1&$0.29\pm0.062$&$-0.050\pm0.011$&$0.29\pm0.062$&$-0.050\pm0.011$\\
 	Gegenbauer-2&$0.31\pm0.071$&$-0.050\pm0.013$&$0.31\pm0.071$&$-0.050\pm0.013$\\
 	QCDSR&$0.29\pm0.061$&$-0.050\pm0.010$&$0.29\pm0.061$&$-0.050\pm0.010$\\
 	Exponential&$0.27\pm0.11$&$-0.045\pm0.017$&$0.27\pm0.11$&$-0.045\pm0.017$\\
 	Free-parton&$0.38\pm0.15$&$-0.063\pm0.024$&$0.38\pm0.15$&$-0.063\pm0.024$\\
 	\hline
 	\hline
 	heavy-LCSR\cite{Wang:2009hra}&$0.023^{+0.006}_{-0.005}$&$-0.039^{+0.009}_{-0.009}$&$0.023^{+0.006}_{-0.005}$&$-0.039^{+0.009}_{-0.009}$\\
 	light-LCSR- $\mathcal{A}$\cite{Khodjamirian:2011jp}&$0.14^{+0.03}_{-0.03}$&$-0.054^{+0.016}_{-0.013}$&$0.14^{+0.03}_{-0.03}$&$-0.028^{+0.012}_{-0.009}$\\
 	light-LCSR- $\mathcal{P}$\cite{Khodjamirian:2011jp}&$0.12^{+0.03}_{-0.04}$&$-0.047^{+0.015}_{-0.013}$&$0.12^{+0.03}_{-0.03}$&$-0.016^{+0.007}_{-0.005}$\\
 	QCD-light-LCSR\cite{Huang:2004vf}&$0.018$&-0.028&0.018&-0.028\\
 	HQET-light-LCSR\cite{Huang:2004vf}&-0.002&-0.015&-0.002&-0.015\\
   PQCD-Exponential\cite{Han:2022srw}&$0.27\pm0.12$&$0.008\pm0.005$&$0.31\pm0.13$&$0.014\pm0.010$\\
   PQCD-Free-parton\cite{Han:2022srw}&$0.24\pm0.10$&$0.007\pm0.004$&$0.27\pm0.16$&$0.014\pm0.008$\\
 	CCQM\cite{Gutsche:2014zna}&0.080&-0.036&0.007&-0.001\\
 	RQM\cite{Faustov:2016pal}&0.169&-0.050&0.196&-0.0002\\
 	LFQM\cite{Wei:2009np}&0.1131&-0.0356&0.1112&-0.0097\\
 	LQCD\cite{Detmold:2015aaa}&$0.22\pm0.08$&$0.04\pm0.12$&$0.12\pm0.14$&$0.04\pm0.31$ \\
 	\hline
 	\hline
 \end{tabular}
 \label{form factors at q0:proton}	
\end{table}
\begin{table}
	\centering
	 \caption{The results of form factors $f_i$ and $g_i$ for $\Lambda_b \rightarrow N^*$ at $q^2=0$ under three interpolating currents for the $N^*$(Ioffe, Tensor and LP) with five different LCDA models of $\Lambda_{b}$ baryon(Gegenbauer-1,Gegenbauer-2, QCDSR, Exponential and Free-parton) in this work. The results for the form factor $F_2(0)$ and $F_3(0)$ are same, $F_2(0)=F_3(0)$, as well as for $G_2(0)$ and $G_3(0)$. For comparison, we also collect the results of the form factors from other works. }
	\begin{tabular}{c c c c c c c}
		\hline
		\hline
		&$F_1$&$F_2$&$G_1$&$G_2$\\
		\hline
		\textbf{Ioffe Current}\\
		Gegenbauer-1&$0.23\pm0.57$&$0.002\pm0.12$&$0.23\pm0.57$&$0.002\pm0.12$\\
		Gegenbauer-2&$0.20\pm0.15$&$0.009\pm0.029$&$0.20\pm0.15$&$0.009\pm0.029$\\
		QCDSR&$0.015\pm0.021$&$0.019\pm0.005$&$0.015\pm0.021$&$0.019\pm0.005$\\
		Exponential&$0.029\pm0.031$&$0.017\pm0.008$&$0.029\pm0.031$&$0.017\pm0.008$\\
		Free-parton&$0.006\pm0.026$&$0.028\pm0.015$&$0.006\pm0.026$&$0.028\pm0.015$\\
		\hline
		\textbf{Tensor Current}\\
		Gegenbauer-1&$0.32\pm0.29$&$-0.11\pm0.087$&$0.32\pm0.29$&$-0.11\pm0.087$\\
		Gegenbauer-2&$0.32\pm0.069$&$-0.10\pm0.023$&$0.32\pm0.069$&$-0.10\pm0.023$\\
		QCDSR&$0.10\pm0.019$&$-0.035\pm0.007$&$0.10\pm0.019$&$-0.035\pm0.007$\\
		Exponential&$0.10\pm0.062$&$-0.036\pm0.021$&$0.10\pm0.062$&$-0.036\pm0.021$&\\
		Free-parton&$0.14\pm0.089$&$-0.050\pm0.032$&$0.14\pm0.089$&$-0.050\pm0.032$\\
		\hline
		\textbf{LP Current}\\
	    Gegenbauer-1&$1.16\pm0.22$&$-0.34\pm0.065$&$1.16\pm0.22$&$-0.34\pm0.065$\\
		Gegenbauer-2&$1.22\pm0.24$&$-0.34\pm0.075$&$1.22\pm0.24$&$-0.34\pm0.075$\\
		QCDSR&$1.16\pm0.22$&$-0.34\pm0.064$&$1.16\pm0.22$&$-0.34\pm0.064$\\
		Exponential&$1.07\pm0.42$&$-0.30\pm0.11$&$1.07\pm0.42$&$-0.30\pm0.11$\\
		Free-parton&$1.48\pm0.60$&$-0.43\pm0.16$&$1.48\pm0.60$&$-0.43\pm0.16$\\
		\hline
		\hline
		LCSR(1)\cite{Emmerich:2016jjm}&$-0.562\pm0.015$&$0.451\pm0.0133$&$0.523\pm0.014$&$-0.454\pm0.013$\\
		LCSR(2)\cite{Emmerich:2016jjm}&$-0.185\pm0.005$&$0.184\pm0.006$&$0.143\pm0.004$&$-0.093\pm0.003$\\
		LCSR-1\cite{Aliev:2019thw}&$-0.297\pm0.080$&$-0.213\pm0.064$&$-0.028\pm0.084$&$0.106\pm0.031$\\
		LCSR-2\cite{Aliev:2019thw}&$-0.202\pm0.060$&$-0.0640\pm0.0018$&$-0.144\pm0.043$&$0.062\pm0.002$\\ 
		\hline
		\hline
	\end{tabular}
\label{form factors at q0:Nstart}	
\end{table}
Some comments on the numerical results of the form factors obtained in  Table.\ref{form factors at q0:proton} and Table.\ref{form factors at q0:Nstart}  are in order:
\begin{itemize}
    \item From the Table.\ref{form factors at q0:proton} and Table.\ref{form factors at q0:Nstart}, we can see that both for the  $\Lambda_b \rightarrow p $ and  $\Lambda_b \rightarrow N^*$ form factors, the form factor relations from heavy quark symmetry in Eq. (\ref{hqlimit}) hold.  This result is natural in the LCSR with $\Lambda_b$-LCDAs, since we have taken the heavy quark limit in the definition of the $\Lambda_b$-LCDAs. To obtain the symmetry breaking effect, one must introduce the power suppressed contribution in HQET, which exceeds the scope of the present study.
 
 \item The results of the form factors based on QCDSR model, Exponential model, Free-parton model are consistent with each other, while the results of the form factors based on Gegenbauer-1 model and Gegenbauer-2 model are significantly larger. This may indicate that Gegenbauer expansion of  the LCDAs of $\Lambda_b$-baryon is not applicable to the $\Lambda_b$-baryon decays, since the light-quarks inside the $\Lambda_b$ baryon are soft quarks, albeit that the evolution equation contains ERBL-like term.  In our following  calculations, we will not take advantage of  the results from the Gegenbauer-1 model and Gegenbauer-2 model.
 
 \item The LP current  can give rise to the unique leading power $\Lambda_b \rightarrow p$ form factor at SCET limit.  The obtained results for the form factors $\xi(0)$ from various LCDAs-models of $\Lambda_b$ baryon read 
 \begin{eqnarray}
 \xi(0)&=&0.223\pm 0.047\,\,\,{\rm (QCDSR\,\, model)},\,\,\,\nonumber \\
 \xi(0)&=&0.277\pm 0.125
 \,\,\,{\rm (Exponential \,\, model)},\nonumber \\
 \,\,\,\xi(0)&=&0.285\pm 0.129\,\,\,{\rm (Free\,\, parton\,\, model)},\,\,\,
 \end{eqnarray}
and these results are close to the form factors shown in Table.\ref{form factors at q0:proton} where the quark field is not replaced by the SCET field. Compared with the results from the Ioffe and tensor current, the results from the LP current are much larger, which implies that the high power correction which has been picked up when the Ioffe current or tensor current being employed  is of great importance. As for the $\Lambda_b \rightarrow N^*$ form factors, the results from the LP current are extraordinarily large because that the normalization constants $\lambda_3^{N^*}$ is about four times smaller than $\lambda_3^{p}$. It is worth noting that it is more difficult to identify the negative parity baryon $N^*$ on lattice than the nucleon. Therefore, the value of  normalization constant $\lambda_3^{N^*}$ maybe is not as reliable as the proton normalization constant.  

\item The form factors for $\Lambda_b \rightarrow p$ from the Ioffe and tensor current are very close to each other once the contribution of the $N^*$ baryon is included in the hadronic dispersion relation. This conclusion is consistent with that in \cite{Khodjamirian:2011jp} where the $\Lambda_b \rightarrow p$ form factors calculated by using LCSR with the LCDAs of the nucleon with two different interpolating currents(the axial-vector and pseudoscalar currents) for the $\Lambda_b$ baryon  are close to each other  under the condition that the contribution of the negative-parity heavy $\Lambda_b^*$ baryon was included. 

\item The form factors for $\Lambda_b \rightarrow N^*$ from  the Ioffe current are much smaller than that from the tensor current because of the large numerical cancellations between  the different invariant amplitudes, specifically speaking, the contribution from the invariant amplitude ${\rm Im}_s \Pi^{Io}_3$ is significantly cancelled by that from  the invariant amplitudes ${\rm Im}_s \Pi^{\rm Io}_4 ,{\rm Im}_s \Pi^{\rm Io}_1, {\rm Im}_s \Pi^{\rm Io}_2 $ in Eq.(\ref{Fi}), when the Ioffe current is employed to calculate the form factor $F_1$ for $\Lambda_b \rightarrow N^*$, and the result is very sensitive to the models of LCDA of $\Lambda_b$-baryon after the cancellation.  The large discrepancy between the predictions from the Ioffe and tensor current indicates that  the form factors for $\Lambda_b \rightarrow N^*$ are very sensitive to the interpolating currents. 
\item It was mentioned in \cite{Ioffe:1981kw} that the Ioffe current coupling to the lowest baryonic state(the proton) is stronger than to the higher state(the $N^*$), then the predicted results of the $\Lambda_b \rightarrow N^*$ form factors are probably  less accurate than that of the $\Lambda_b \rightarrow p$ form factors. We hope the future experimental observables of the semi-leptonic decay  $\Lambda_b\rightarrow N^*l\nu$ can help us to determine which kind of the interpolating current is more preferable.
\end{itemize}

For comparison, we also collect the predictions on the $\Lambda_b \rightarrow p$ and $\Lambda_b \rightarrow N^*$ form factors in other researches in Table.\ref{form factors at q0:proton} and Table.\ref{form factors at q0:Nstart} respectively. The $\Lambda_b \rightarrow p$ form factors was studied  in \cite{Wang:2009hra} based on LCSR with $\Lambda_b$-LCDAs  where  the Gegenbauer-1 model of $\Lambda_b$-LCDAs  was employed and the nucleon(without the negative-parity) was interpolated by the CZ current, and the predicted  from factors $f_1(0)$ and $g_1(0)$ are about an order of magnitude smaller than our results. This difference arises from the different model of LCDAs, the different interpolating currents and the absence of the negative-parity partner of the nucleon. The results in \cite{Khodjamirian:2011jp} are consistent with our results, which implies that the results of  light-hadron LCSR and heavy-hadron LCSR can give rise to the consistent results. In \cite{Huang:2004vf}, the LCSR with the nucleon-distribution amplitudes was employed where the $\Lambda_b$ was also interpolated by CZ current and only the ground state  was considered in the hadronic dispersion integral, and the results of the $\Lambda_b \rightarrow p$ form factors are also about an order of magnitude smaller than ours. The predictions of $\Lambda_b \rightarrow p$ form factors from  the  covariant constituent quark model\cite{Gutsche:2014zna}, the relativistic quark model  \cite{Faustov:2016pal}, the light-front quark model\cite{Wei:2009np}, and the lattice simulation\cite{Detmold:2015aaa} are also listed in  Table.\ref{form factors at q0:proton}. It can be seen that our results are consistent with these predictions by considering the theoretical uncertainty. The $\Lambda_b \rightarrow N^*$ form factors were studied in  \cite{Emmerich:2016jjm} where the LCSR with $N^*$-LCDA was employed.   The interpolating current of  $\Lambda_b$ baryon was adopted as the axial-vector current and two different models of the LCDAs of $N^*$  were used in the calculation. The results from these two models( called LCSR(1) and LCSR(2) respectively) have large discrepancy   because of  the different parameters of twist-4 LCDA  in these two models. In \cite{Aliev:2019thw}, these form factors were revisited where the $\Lambda_b$ baryon was interpolated by the most general current with an arbitrary parameter $\beta$ for the mixing of different components   and the contribution of negative-parity $\Lambda_b^*$ baryon was included, and the results still significantly depend on the models of the LCDAs, and they are not consistent with our predictions. Thus we hope our studies can provide useful hints on the  LCDAs of the negative-parity states. Recently, a new study about the $\Lambda_b \rightarrow p$ form factors based on the perturbation QCD(PQCD) approach was presented in \cite{Han:2022srw}. Different from the previous PQCD calculations\cite{Lu:2009cm, Shih:1998pb}, this work includes the contribution from the higher twist LCDAs of $\Lambda_b$ baryon and proton, and the results indicate that the higher twist LCDAs give the dominant contributions since the important endpoint region of the convolution of the LCDAs and the hard kernels has been picked up. The renewed numerical result from the PQCD calculation can also be consistent with our predictions.

To guarantee the light-cone dominance, the LCSR predictions for the form factors  are reliable up to a limited region of the momentum transfer, namely $q^2\leq 10{\rm GeV^2}$. In order to extend the LCSR predictions  to the whole physical region, we take advantage of the z-series parameterization in the BCL-version suggested in \cite{Bourrely:2008za}, 
\begin{eqnarray} 
 z(q^2,t_0)=\frac{\sqrt{t_+-q^2}-\sqrt{t_+-t_0}}{\sqrt{t_+-q^2}+\sqrt{t_+-t_0}}  ,
\end{eqnarray}
where $t_{\pm}=(m_{\Lambda_b}\pm m_{p(N^*)})^2$, and $t_0=t_+-\sqrt{t_+-t_{-}}\sqrt{t_+-t_{min}}$ is chosen to reduce the interval of $z$ after mapping $q^2$ to $z$ with the interval $t_{min}<q^2<t_{-}$. In the numerical analysis, we adopt $t_{min}=-6{\rm GeV^2}$. To achieve the best parametrization of the form factors, we employ the following parametrization
\begin{eqnarray}
   f(q^2)=\frac{f_i(0)}{1-q^2/(m_{pole}^f)^2} \Big\{1+a_1(z(q^2,t_0)-z(0,t_0))\Big\} ,
\end{eqnarray}
For the pole masses, we adopt $m_{pole}=m_{B^*}=5.325$ GeV for the from factors $f_1,f_2, F_1,F_2$; $m_{pole}=m_{B_1}=5.723$ GeV for the from factors $g_1,g_2, G_1,G_2$; $m_{pole}=m_{B_0}=5.749$ GeV for the from factors $f_3, F_3$; $m_{pole}=m_{B}=5.280$ GeV for the from factors $g_3, G_3$.


 \begin{table}
 	  \scriptsize
       \centering
       \caption{The $z$-fit parameters, namely $f(0)$ and $a_1$,as well as the correlation coefficient $\rho$ between them obtained from LCSR for $\Lambda_b  \rightarrow p$  under two interpolating current (Ioffe and tensor) with  three different LCDA models of $\Lambda_b$ baryon (QCDSR, Exponential and Free-parton).} 
       \begin{tabular}{c c c c c c c c} 
       \hline 
       \hline
       &&$f_1$&$f_2$&$f_3$&$g_1$&$g_2$&$g_3$ \\ 
       \hline
       \multicolumn{8}{c}{\textbf{Ioffe Current}}\\
       \hline
       \multirow{3}{*}{QCDSR}&f(0)&$0.13\pm0.020$&$-0.021\pm0.004$&$-0.021\pm0.004$&$0.13\pm0.020$&$-0.020\pm0.004$&$-0.021\pm0.004$\\
       &$a_1$&$-8.73\pm0.83$&$-13.80\pm0.98$&$-14.47\pm1.01$&$-9.49\pm0.84$&$-14.44\pm1.00$&$-13.72\pm0.97$\\
       &$\rho$&$0.091$&$-0.060$&$-0.062$&$0.089$&$-0.060$&$-0.062$\\
       \hline
        \multirow{3}{*}{Exponential}&f(0)&$0.15\pm0.055$&$-0.026\pm0.012$&$-0.026\pm0.012$&$0.15\pm0.057$&$-0.026\pm0.012$&$-0.026\pm0.012$\\
       &$a_1$&$-7.49\pm2.17$&$-12.65\pm2.41$&$-13.42\pm2.45$&$-8.30\pm2.16$&$-13.38\pm2.45$&$-12.56\pm2.41$\\
       &$\rho$&$0.362$&$-0.351$&$-0.353$&$0.359$&$-0.352$&$-0.350$\\
       \hline
       \multirow{3}{*}{Free-parton}&f(0)&$0.17\pm0.072$&$-0.028\pm0.015$&$-0.028\pm0.015$&$0.17\pm0.073$&$-0.028\pm0.015$&$-0.028\pm0.015$ \\
       &$a_1$&$-9.06\pm2.25$&$-14.39\pm2.80$&$-15.11\pm2.89$&$-9.84\pm2.54$&$-15.06\pm2.88$&$-14.31\pm2.79$\\
       &$\rho$&$0.232$&$-0.236$&$-0.238$&$0.231$&$-0.238$&$-0.235$\\
       \hline
       \multicolumn{8}{c}{\textbf{Tensor Current}}\\
       \hline
        \multirow{3}{*}{QCDSR}&f(0)&$0.11\pm0.019$&$-0.021\pm0.004$&$-0.021\pm0.004$&$0.11\pm0.019$&$-0.021\pm0.004$&$-0.021\pm0.004$\\
       &$a_1$&$-8.56\pm0.86$&$-13.59\pm1.00$&$-14.26\pm1.03$&$-9.33\pm0.87$&$-14.22\pm1.03$&$-13.51\pm1.00$\\
       &$\rho$&$0.071$&$-0.041$&$-0.040$&$0.068$&$-0.039$&$-0.041$\\
       \hline
       \multirow{3}{*}{Exponential}&f(0)&$0.12\pm0.044$&$-0.025\pm0.009$&$-0.024\pm0.009$&$0.12\pm0.045$&$-0.024\pm0.009$&$-0.025\pm0.009$\\
       &$a_1$&$-7.35\pm2.05$&$-12.66\pm2.07$&$-13.43\pm2.11$&$-8.16\pm2.05$&$-13.39\pm2.10$&$-12.56\pm2.07$\\
       &$\rho$&$0.337$&$-0.368$&$-0.370$&$0.334$&$-0.370$&$-0.368$\\
       \hline
       \multirow{3}{*}{Free-parton}&f(0)&$0.16\pm0.067$&$-0.031\pm0.014$&$-0.031\pm0.014$&$0.16\pm0.069$&$-0.031\pm0.014$&$-0.031\pm0.014$\\
       &$a_1$&$-8.82\pm2.32$&$-14.08\pm2.44$&$-14.82\pm2.51$&$-9.61\pm2.32$&$-14.78\pm2.50$&$-13.99\pm2.43$\\
       &$\rho$&$0.293$&$-0.329$&$-0.333$&$0.292$&$-0.332$&$-0.329$\\
       \hline
       \hline
       \end{tabular}
\label{fit:proton}
\end{table}

\begin{table}
	\scriptsize
	\centering
	\caption{The $z$-fit parameters, namely $f(0)$ and $a_1$,as well as the correlation coefficient $\rho$ between them obtained from LCSR for $\Lambda_b  \rightarrow N^*$  under two interpolating current (Ioffe and tensor) with three different LCDA models of $\Lambda_b$ baryon (QCDSR, Exponential and Free-parton).} 
	\begin{tabular}{c c c c c c c c} 
		\hline 
		\hline
		&&$f_1$&$f_2$&$f_3$&$g_1$&$g_2$&$g_3$ \\ 
		\hline
		\multicolumn{8}{c}{\textbf{Ioffe Current}}\\
		\hline
		\multirow{3}{*}{QCDSR}&f(0)&$0.016\pm0.011$&$0.018\pm0.004$&$0.018\pm0.004$&$0.017\pm0.011$&$0.018\pm0.004$&$0.018\pm0.004$\\
		&$a_1$&$-14.76\pm11.13$&$-15.65\pm1.31$&$-16.42\pm1.35$&$-15.59\pm11.02$&$-16.38\pm1.35$&$-15.55\pm1.31$\\
        &$\rho$&$0.246$&$-0.011$&$-0.014$&$0.239$&$-0.013$&$-0.011$\\
		\hline
		\multirow{3}{*}{Exponential}&f(0)&$0.029\pm0.022$&$0.017\pm0.005$&$0.017\pm0.005$&$0.031\pm0.023$&$0.017\pm0.005$&$0.017\pm0.005$\\
		&$a_1$&$-11.13\pm5.16$&$-14.85\pm2.01$&$-15.74\pm2.05$&$-12.07\pm5.19$&$-15.69\pm2.05$&$-14.74\pm2.00$\\
        &$\rho$&$0.479$&$0.328$&$0.330$&$0.478$&$0.329$&$0.328$\\
		\hline
		\multirow{3}{*}{Free-parton}&f(0)&$0.007\pm0.015$&$0.028\pm0.011$&$0.028\pm0.011$&$0.007\pm0.015$&$0.028\pm0.011$&$0.028\pm0.011$\\
		&$a_1$&$-28.96\pm51.95$&$-15.98\pm2.66$&$-16.85\pm2.71$&$-29.54\pm51.15$&$-16.80\pm2.71$&$-15.87\pm2.66$\\
        &$\rho$&$0.681$&$0.388$&$0.390$&$0.675$&$0.389$&$0.387$\\
		\hline
		\multicolumn{8}{c}{\textbf{Tensor Current}}\\
		\hline
		\multirow{3}{*}{QCDSR}&f(0)&$0.10\pm0.016$&$-0.033\pm0.007$&$-0.033\pm0.007$&$0.10\pm0.017$&$-0.033\pm0.007$&$-0.033\pm0.007$\\
		&$a_1$&$-9.70\pm1.01$&$-15.99\pm1.19$&$-16.78\pm1.23.$&$-10.61\pm1.02$&$-16.73\pm1.23$&$-15.90\pm1.19$\\
        &$\rho$&$0.079$&$-0.041$&$-0.390$&$0.076$&$-0.038$&$-0.041$\\
		\hline
		\multirow{3}{*}{Exponential}&f(0)&$0.11\pm0.040$&$-0.037\pm0.014$&$-0.037\pm0.014$&$0.11\pm0.040$&$-0.037\pm0.014$&$-0.037\pm0.014$\\
		&$a_1$&$-8.28\pm2.43$&$-14.93\pm2.46$&$-15.84\pm2.51$&$-9.24\pm2.42$&$-15.79\pm2.50$&$-14.82\pm2.45$\\
        &$\rho$&$0.332$&$-0.368$&$-0.371$&$0.330$&$-0.374$&$0.365$\\
		\hline
		\multirow{3}{*}{Free-parton}&f(0)&$0.14\pm0.058$&$-0.048\pm0.022$&$-0.048\pm0.022$&$0.14\pm0.02$&$-0.048\pm0.022$&$-0.048\pm0.022$\\
		&$a_1$&$-9.99\pm2.73$&$-16.59\pm2.91$&$-17.46\pm2.99$&$-10.93\pm2.74$&$-17.42\pm2.99$&$-16.49\pm2.89$\\
        &$\rho$&$0.291$&$-0.330$&$-0.334$&$0.288$&$-0.333$&$0.329$\\
		\hline
		\hline
	\end{tabular}
\label{fit:Nstart}
\end{table}

\begin{figure}
	\begin{center}
		\includegraphics[width=1\columnwidth]{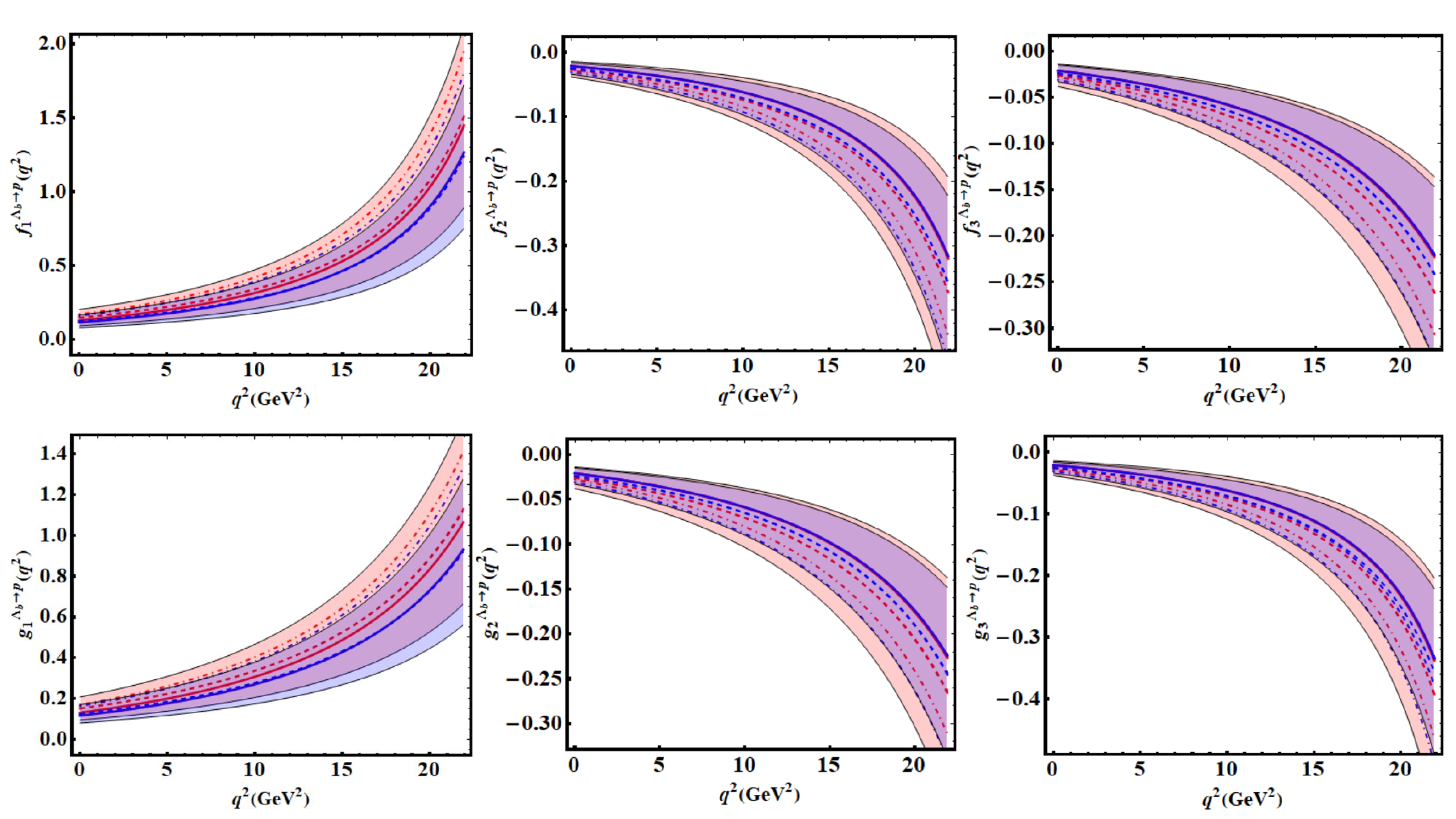}
		\caption{$\Lambda_b\rightarrow p$ transition form factor obtained from LCSR at $q^2\leq 10$ GeV$^2$ and extrapolation to large $q^2$ using the $z$ serizes-parametrization: the red(blue) solid line correspond to the Ioffe(tensor) current for proton with QCDSR-LCDA model for $\Lambda_b$, the red(blue) dashed line correspond to the Ioffe(tensor) current for pronto with Exponential LCDA model for $\Lambda_b$ and the red(blue) dot-dashed line correspond to the Ioffe(tensor) current for proton with Free-parton LCDA model for $\Lambda_b$. With the LCDA of $\Lambda_b$ being fixed as the Exponential model, the red(blue) band correspond to the range of the uncertainties of the form factors under the Ioffe(tensor) interpolating current.}
		\label{fig:q dependence for proton}
	\end{center}
\end{figure}

\begin{figure}
	\begin{center}
		\includegraphics[width=1\columnwidth]{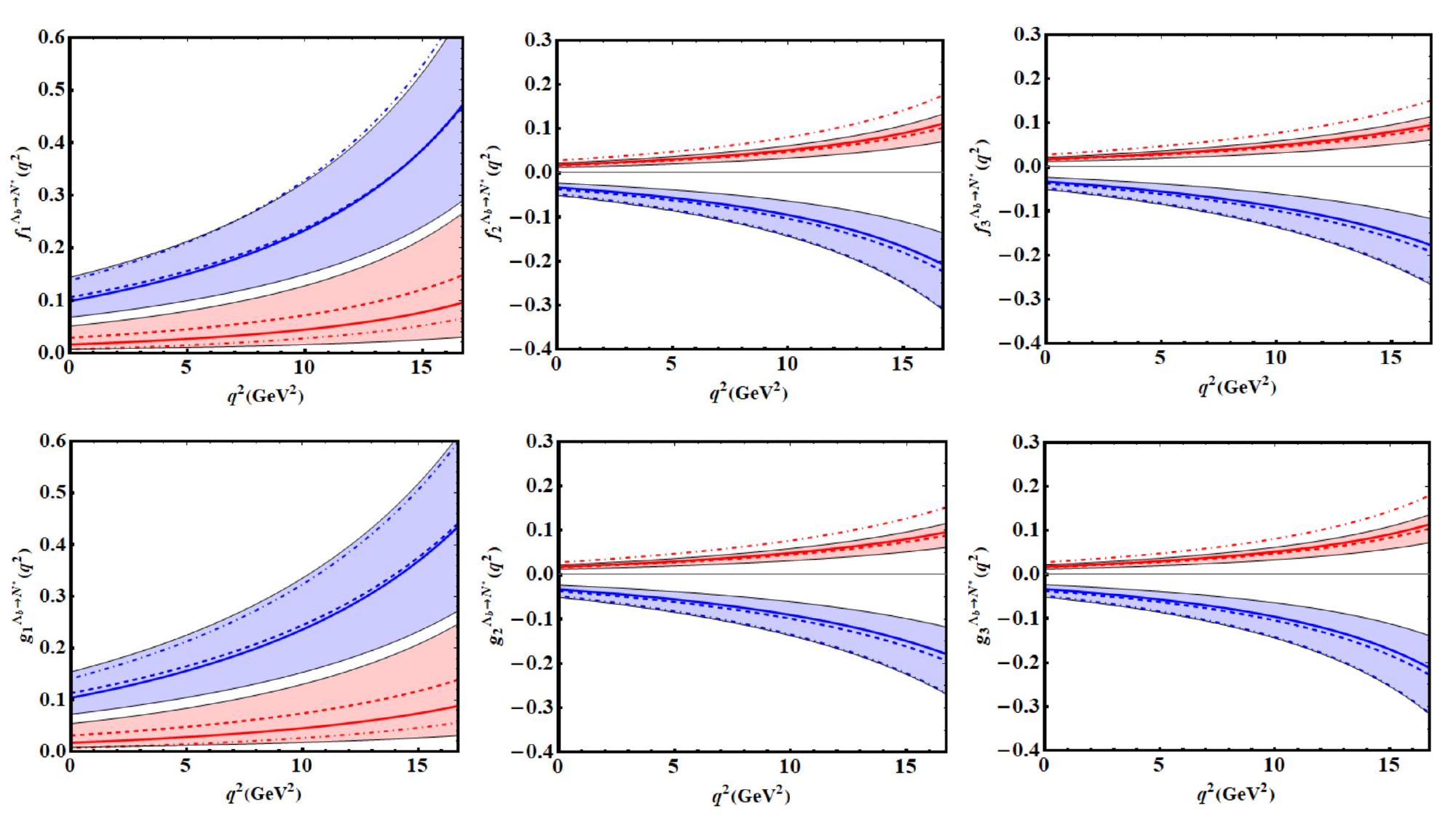}
		\caption{Same as Fig. \ref{fig:q dependence for proton} but for $\Lambda_b\rightarrow N^*$ transition form factor.}
		\label{fig:q dependence for Nstart}
	\end{center}
\end{figure}

To determine the central values, the uncertainties ,and the correlation coefficients of the z-fit parameters $f_{i}(0)$ and $a_{1i}$ for each $\Lambda_b \rightarrow p,\ N^*$form factor, we first generate an ensemble of correlated LCSR data points of these form factors. The LCSR data points of these form factors are calculated at $q^2=\{-6,-3,0,3,6\}$ Gev$^2$ with $N=500$ ensembles of the input parameter set (including $M^2$, $s_0$,and the parameters of LCDAs for $\Lambda_b$) where the value of the input parameters are randomly distributed according to a multivariate join distribution\cite{SentitemsuImsong:2014plu}. Then we fit the $z$-series expansion to all the $\Lambda_b \rightarrow p, N^*$ form factors to get the $z$-fit parameters $f_{i}(0)$ and $a_{1i}$, as weel as the correlation coefficients between them. The fitting results are given in Table.\ref{fit:proton} and Table.\ref{fit:Nstart} respectively, and the $q^2$  dependence of these form factors are shown in Fig.\ref{fig:q dependence for proton} and Fig.\ref{fig:q dependence for Nstart} respectively, where both  the Ioffe current and tensor current are employed to interpolate the baryon states, and  the QCDSR model, the Exponential model and  the Free-parton model of  $\Lambda_b$ baryon  LCDA are all taken into account for comparison. In oreder to test whether the results form these interpolating currents and LCDAs models are consistent within uncertainties, we also display the error bands of the form factors under two different interpolating currents with Exponential LCDA model for $\Lambda_b$.  From  Fig.\ref{fig:q dependence for proton}, we can see that the form factors of $\Lambda_b \rightarrow p$ are insensitive to the interpolating currents apparently and the results form the different interpolating currents and LCDAs models are consistent within the uncertainties. The predictions based on QCDSR model are very close to that based on Exponential model, and the Free-parton model give rise to little larger values of the form factors.  As for the form factors of $\Lambda_b \rightarrow N^*$, the results are sensitive to the interpolating currents and the results from the different LCDAs models of $\Lambda_b$ are consistent within uncertainties. We note that the uncertainties of the fitted results are mainly from the LCDAs of the $\Lambda_b$ baryon. Therefore, it is an essential task to determine the most preferable model and to constrain the parameters of the LCDAs of $\Lambda_b$ baryon in the study on the heavy baryon decays.

\subsection{Semi-leptonic decays of $\Lambda_b \rightarrow p(N^*)l^-\bar{\nu_l}$}

Now, we can take advantage of the obtained $\Lambda_b \rightarrow p(N^*)$ form factors to calculate the branching ratios of the semileptonic decays, as well as different asymmetry parameters and the other observables.    In doing this, we employ the Ioffe current and the tensor current and three suitable LCDA models of $\Lambda_{b}$-baryon which are QCDSR model, Exponential model and Free-parton model to evaluate the correlation function used in the LCSRs.  We hope the future experiments will measure these observables so that it can help us to make certain which kind of the interpolating currents and which type of LCDA models of $\Lambda_b$ baryon are correct.

\begin{table}
	\centering
	\caption{The predictions for the branching fractions, the averaged leptonic forward-backward asymmetry$\langle A_{FB}\rangle$, the averaged final hadronic polarization $\langle P_{B}\rangle$ and the averaged lepton polarization $\langle P_{l}\rangle$ for $\Lambda_b\rightarrow pl^-\bar{\nu}_l$ under two interpolating current (Ioffe and Tensor) with  three different LCDA models of $\Lambda_b$ baryon (QCDSR, Exponential and Free-parton). For comparison, we also collect the results of these observables from other works and the experiment.}
	\begin{tabular}{c c c c c c}
		\hline
		\hline
		Model&$l$&Br($\times 10^{-4}$)&$\langle A_{FB}\rangle$&$\langle P_B\rangle$&$\langle P_l\rangle$\\
		\hline
		\multicolumn{6}{c}{\textbf{Ioffe current}}\\
		\hline
		\multirow{3}{*}{QCDSR}&e&$3.74\pm0.92$&$0.33\pm0.01$&$-0.95\pm0.05$&$-1.00\pm0.00$\\
		&$\mu$&$3.73\pm0.91$&$0.32\pm0.01$&$-0.95\pm0.05$&$-0.99\pm0.00$\\
		&$\tau$&$2.59\pm0.63$&$0.15\pm0.01$&$-0.93\pm0.04$&$-0.55\pm0.04$\\
		\hline
		\multirow{3}{*}{Exponential}&e&$4.28\pm2.40$&$0.33\pm0.03$&$-0.96\pm0.09$&$-1.00\pm0.00$\\
		&$\mu$&$4.27\pm2.40$&$0.32\pm0.03$&$-0.96\pm0.09$&$-0.99\pm0.00$\\
		&$\tau$&$2.93\pm1.63$&$0.15\pm0.03$&$-0.95\pm0.08$&$-0.54\pm0.09$\\
		\hline
		\multirow{3}{*}{Free-parton}&e&$6.73\pm4.38$&$0.33\pm0.03$&$-0.96\pm0.12$&$-1.00\pm0.00$\\
		&$\mu$&$6.72\pm4.37$&$0.33\pm0.03$&$-0.96\pm0.12$&$-0.99\pm0.00$\\
		&$\tau$&$4.69\pm3.03$&$0.15\pm0.03$&$-0.94\pm0.10$&$-0.55\pm0.11$\\
		\hline
		\multicolumn{6}{c}{\textbf{Tensor current}}\\
		\hline
		\multirow{3}{*}{QCDSR}&e&$2.82\pm0.75$&$0.33\pm0.01$&$-0.96\pm0.04$&$-1.00\pm0.00$\\
		&$\mu$&$2.81\pm0.74$&$0.33\pm0.01$&$-0.96\pm0.04$&$-0.99\pm0.00$\\
		&$\tau$&$1.95\pm0.52$&$0.15\pm0.01$&$-0.95\pm0.03$&$-0.54\pm0.04$\\
		\hline
		\multirow{3}{*}{Exponential}&e&$2.86\pm1.54$&$0.33\pm0.02$&$-0.97\pm0.07$&$-1.00\pm0.00$\\
		&$\mu$&$2.85\pm1.54$&$0.33\pm0.02$&$-0.97\pm0.07$&$-0.99\pm0.00$\\
		&$\tau$&$1.95\pm1.05$&$0.15\pm0.03$&$-0.96\pm0.06$&$-0.54\pm0.08$\\
		\hline
		\multirow{3}{*}{Free-parton}&e&$5.69\pm3.64$&$0.34\pm0.02$&$-0.97\pm0.09$&$-1.00\pm0.00$\\
		&$\mu$&$5.68\pm3.64$&$0.34\pm0.02$&$-0.97\pm0.09$&$-0.99\pm0.00$\\
		&$\tau$&$3.95\pm 2.51$&$0.15\pm0.03$&$-0.96\pm0.07$&$-0.54\pm0.10$\\
		\hline
		\hline
		\multirow{3}{*}{RQM\cite{Faustov:2016pal}}&e&4.5&0.346&-&-0.91 \\
		&$\mu$&4.5&0.344&-&-0.91\\
		&$\tau$&2.9&0.185&-&-0.89\\
		\hline
		\multirow{2}{*}{LQCD\cite{Dutta:2015ueb}}&e($\mu$)&3.89&-&-&-\\
		&$\tau$&2.75&-&-&-\\
		\hline
		light-LCSR\cite{Khodjamirian:2011jp}&$e(\mu)$&$4.0^{+2.3}_{-2.0}$&-&-&-\\
		\hline
		Exp.\cite{LHCb:2015eia}&$\mu$&$4.1\pm1.0$ &-&-&-\\ 
		\hline
		\hline
	\end{tabular}
	\label{observable:proton}
\end{table}
\begin{table}[h]
	\centering
	\caption{The predictions for the branching fractions, the averaged leptonic forward-backward asymmetry$\langle A_{FB}\rangle$, the averaged final hadron polarization $\langle P_{B}\rangle$ and the averaged lepton polarization $\langle P_{l}\rangle$ for $\Lambda_b\rightarrow N^*l^-\bar{\nu}_l$ under two interpolating current(Ioffe and tensor) with three different LCDA models of $\Lambda_b$ baryon(QCDSR, Exponential and Free-parton).}
	\begin{tabular}{c c c c c c}
		\hline
		\hline
		Model&$l$&Br($\times 10^{-5}$)&$\langle A_{FB}\rangle$&$\langle P_B\rangle$&$\langle P_l\rangle$\\
		\hline
		\multicolumn{6}{c}{\textbf{Ioffe current}}\\
		\hline
		\multirow{3}{*}{QCDSR}&e&$0.88\pm0.69$&$-0.06\pm0.12$&$-0.13\pm0.32$&$-1.00\pm0.00$\\
		&$\mu$&$0.88\pm0.68$&$-0.06\pm0.12$&$-0.13\pm0.32$&$-0.99\pm0.00$\\
		&$\tau$&$0.55\pm0.40$&$-0.10\pm0.06$&$-0.12\pm0.27$&$-0.57\pm0.18$\\
		\hline
		\multirow{3}{*}{Exponential}&e&$1.55\pm1.44$&$0.04\pm0.13$&$-0.37\pm0.39$&$-1.00\pm0.00$\\
		&$\mu$&$1.55\pm1.44$&$0.04\pm0.13$&$-0.37\pm0.39$&$-0.99\pm0.00$\\
		&$\tau$&$0.94\pm0.81$&$-0.06\pm0.06$&$-0.34\pm0.31$&$-0.53\pm0.22$\\
		\hline
		\multirow{3}{*}{Free-parton}&e&$0.90\pm1.08$&$-0.25\pm0.24$&$0.30\pm0.49$&$-1.00\pm0.00$\\
		&$\mu$&$0.90\pm1.07$&$-0.25\pm0.24$&$0.30\pm0.49$&$-0.99\pm0.00$\\
		&$\tau$&$0.57\pm0.63$&$-0.19\pm0.14$&$0.25\pm0.44$&$-0.62\pm0.26$\\
		\hline
		\multicolumn{6}{c}{\textbf{Tensor current}}\\
		\hline
		\multirow{3}{*}{QCDSR}&e&$7.86\pm2.22$&$0.28\pm0.01$&$-0.95\pm0.04$&$-1.00\pm0.00$\\
		&$\mu$&$7.85\pm2.21$&$0.27\pm0.01$&$-0.95\pm0.04$&$-0.99\pm0.00$\\
		&$\tau$&$4.17\pm1.22$&$0.07\pm0.02$&$-0.93\pm0.05$&$-0.44\pm0.06$\\
		\hline
		\multirow{3}{*}{Exponential}&e&$8.07\pm4.68$&$0.28\pm0.02$&$-0.96\pm0.06$&$-1.00\pm0.00$\\
		&$\mu$&$8.06\pm4.67$&$0.27\pm0.02$&$-0.96\pm0.06$&$-0.99\pm0.00$\\
		&$\tau$&$4.18\pm2.51$&$0.06\pm0.05$&$-0.95\pm0.08$&$-0.43\pm0.12$\\
		\hline
		\multirow{3}{*}{Free-parton}&e&$15.31\pm10.65$&$0.28\pm0.03$&$-0.96\pm0.09$&$-1.00\pm0.00$\\
		&$\mu$&$15.28\pm10.63$&$0.27\pm0.03$&$-0.96\pm0.09$&$-0.05\pm0.00$\\
		&$\tau$&$8.11\pm5.83$&$0.06\pm0.05$&$-0.94\pm0.11$&$-0.44\pm0.15$\\
		\hline
		\hline
	\end{tabular}
	\label{observable:Nstart}
\end{table}

In order to calculate these observables, it is convenient to introduce the helicity amplitudes\cite{Azizi:2019tcn,Dutta:2018zqp}. The helicity amplitudes can be defined by 
\begin{eqnarray}
	H^{V,A}_{\lambda_{p(N^*)},\lambda_{W^-} }=\epsilon^{\dagger\mu}(\lambda_{W^-})\langle p(N^*),\lambda_{p(N^*)}|V(A)|\Lambda_b, \lambda_{\Lambda_b}\rangle ,
\end{eqnarray}
where the $\lambda_{\Lambda_b}, \lambda_{p(N^*)}, \lambda_{W^-}$ denote the helicity of the $\Lambda_b$ baryon, the proton($N^*$) and the off-shell $W^-$ respectively. The helicity amplitudes $H^{V,A}_{\lambda_{p(N^*)},\lambda_{W^-}}$ can be expressed as functions of the form factors\cite{Zwicky:2013eda,Li:2021qod}

\begin{eqnarray}
	H^{V}_{\frac{1}{2},0}(1/2^+\rightarrow 1/2^{\pm})&=&\frac{\sqrt{Q_{\mp}}}{\sqrt{q^2}}\Big(M_{\pm}f(F)_1(q^2)\mp \frac{q^2}{m_{\Lambda_b}}f(F)_2(q^2)\Big), \nonumber    \\ 
	H^{A}_{\frac{1}{2},0}(1/2^+\rightarrow 1/2^{\pm})&=&\frac{\sqrt{Q_{\pm}}}{\sqrt{q^2}}\Big(M_{\mp}g(G)_1(q^2)\pm \frac{q^2}{m_{\Lambda_b}}g(G)_2(q^2)\Big) , \nonumber \\
	H^{V}_{\frac{1}{2},1}(1/2^+\rightarrow
	1/2^{\pm})&=&\sqrt{2Q_{\mp}}\Big(f(F)_1(q^2)\mp \frac{M_{\pm}}{m_{\Lambda_b}}f(F)_2(q^2)\Big) , \nonumber \\
	H^{A}_{\frac{1}{2},1}(1/2^+\rightarrow
	1/2^{\pm})&=&\sqrt{2Q_{\pm}}\Big(g(G)_1(q^2)\pm \frac{M_{\mp}}{m_{\Lambda_b}}g(G)_2(q^2)\Big) , \nonumber \\
	H^{V}_{\frac{1}{2},t}(1/2^+\rightarrow 1/2^{\pm})&=&\frac{\sqrt{Q_{\pm}}}{\sqrt{q^2}}\Big(M_{\mp}f(F)_1(q^2)\pm \frac{q^2}{m_{\Lambda_b}}f(F)_3(q^2)\Big), \nonumber    \\
	H^{A}_{\frac{1}{2},t}(1/2^+\rightarrow 1/2^{\pm})&=&\frac{\sqrt{Q_{\mp}}}{\sqrt{q^2}}\Big(M_{\pm}g(G)_1(q^2)\mp \frac{q^2}{m_{\Lambda_b}}g(G)_3(q^2)\Big) . 
\end{eqnarray}
where $Q_{\pm}$ is defined as $Q_{\pm}=(m_{\Lambda_b}\pm m_{p(N^*)})^2-q^2$ and $M_{\pm}=m_{\Lambda_b} \pm m_{p(N^*)}$. The negative helicity amplitudes for the involved final baryon with $J^P=1/2^+$ and $1/2^-$ can be obtained by using the relation 
\begin{eqnarray}
	H^V_{-\lambda_{p(N^*)},-\lambda_{W^-}}=\mathcal{P}^V H^V_{\lambda_{p(N^*)},\lambda_{W^-}} , \ \ \ \ 	H^A_{-\lambda_{p(N^*)},-\lambda_{W^-}}=\mathcal{P}^A H^A_{\lambda_{p(N^*)},\lambda_{W^-}} ,
\end{eqnarray} 
with $(\mathcal{P}^V,\mathcal{P}^A)=(+,-)$ and  $(\mathcal{P}^V,\mathcal{P}^A)=(-,+)$, respectively. The total helicity amplitudes for the $V-A$ current are given by 
\begin{eqnarray}
	H_{\lambda_{p(N^*)},\lambda_{W^-}}=H^V_{\lambda_{p(N^*)},\lambda_{W^-}}-H^A_{\lambda_{p(N^*)},\lambda_{W^-}},
\end{eqnarray}

With the helicity amplitudes, the differential angular distribution for the decay $\Lambda_b \rightarrow p(N^*)l^-\bar{\nu}_l$ can be obtained 
\begin{eqnarray}
	\frac{d\Gamma(\Lambda_b \rightarrow p(N^*)l^-\bar{\nu}_l)}{dq^2d\ cos\theta_l}=\frac{G_F^2|V_{ub}|^2q^2|\vec{p}_{p(N^*)}|}{512\pi^3m^2_{\Lambda_b}}\Big(1-\frac{m^2_l}{q^2}\Big)^2\times\Big[A_1+\frac{m_l^2}{q^2}A_2\Big],
\end{eqnarray}
where $G_F$ is the Fermi constant, $V_{ub}$ is the CKM matrix element , $m_l$ is the lepton mass($l=e,\mu ,\tau$), $\theta_l$ is the angle between the three-momentum  of the final $p(N^*)$ baryon and the lepton in the $q^2$ rest frame and 
\begin{eqnarray}
	A_1&=&2\sin^2\theta_l(H^2_{1/2,0}+H^2_{-1/2,0})+(1-\cos \theta_l)^2H^2_{1/2,1}+(1+\cos\theta_l)^2H^2_{-1/2,-1} ,\nonumber \\
	A_2&=&2\cos^2\theta_l(H^2_{1/2,0}+H^2_{-1/2,0})+\sin^2 \theta_l(H^2_{1/2,1}+H^2_{-1/2,-1})+2(H^2_{1/2,t}+H^2_{-1/2,t}) \nonumber \\
	&\ &-4\cos\theta_l(H_{1/2,t}H_{1/2,0}+H_{-1/2,t}H_{-1/2,0}), \nonumber \\
	|\vec{p}_{p(N^*)}|&=&\frac{\sqrt{m_{\Lambda_b}^4+m^4_{p(N^*)}+q^4-2(m^2_{\Lambda_b}m^2_{p(N^*)}+m^2_{p(N^*)}q^2+m^2_{\Lambda_b}q^2)}}{2m_{\Lambda_b}} ,
\end{eqnarray}
The differential decay rate can be obtained by integrating out cos$\theta_l$ 
\begin{eqnarray}
	\frac{d\Gamma(\Lambda_b\rightarrow p(N^*)l^-\bar{\nu}_l)}{dq^2}=\int^1_{-1}\frac{d\Gamma(\Lambda_b \rightarrow p(N^*)l^-\bar{\nu}_l)}{dq^2d\ cos\theta_l}d\cos\theta_l , 
\end{eqnarray}

Additionally, many other important physical observables, e.g., the leptonic forward-backward asymmetry ($A_{FB}$), the final hadron polarization($P_B$) and the lepton polarization($P_l$),  can be expressed in terms of the helicity amplitudes. They are defined as 
\begin{eqnarray}
	A_{FB}(q^2)&=&\frac{\int^1_0 \frac{d\Gamma}{dq^2dcos\theta_l}d\cos\theta_l-\int^0_{-1} \frac{d\Gamma}{dq^2dcos\theta_l}d\cos\theta_l}{\int^1_0 \frac{d\Gamma}{dq^2dcos\theta_l}d\cos\theta_l+\int^0_{-1} \frac{d\Gamma}{dq^2dcos\theta_l}d\cos\theta_l} , \\
	P_B(q^2)&=&\frac{d\Gamma^{\lambda_{p(N^*)}=1/2}/dq^2-d\Gamma^{\lambda_{p(N^*)}=-1/2}/dq^2}{d\Gamma/dq^2} , \\
	P_l(q^2)&=&\frac{d\Gamma^{\lambda_l=1/2}/dq^2-d\Gamma^{\lambda_l=-1/2}/dq^2}{d\Gamma/dq^2}.
\end{eqnarray}
respectively, where 
\begin{eqnarray}
	\frac{d\Gamma^{\lambda_{p(N^*)}=1/2}}{dq^2}&=&\frac{4m^2_l}{3q^2}\Big(H^2_{1/2,1}+H^2_{1/2,0}+3H^2_{1/2,t}\Big)+\frac{8}{3}\Big(H^2_{1/2,0}+H^2_{1/2,1}\Big) , \\
	\frac{d\Gamma^{\lambda_{p(N^*)}=-1/2}}{dq^2}&=&\frac{4m^2_l}{3q^2}\Big(H^2_{-1/2,-1}+H^2_{-1/2,0}+3H^2_{-1/2,t}\Big)+\frac{8}{3}\Big(H^2_{-1/2,0}+H^2_{-1/2,-1}\Big) , \\		\frac{d\Gamma^{\lambda_l=1/2}}{dq^2}&=&\frac{m^2_l}{q^2}\Big[\frac{4}{3}\Big(H^2_{1/2,1}+H^2_{1/2,0}+H^2_{-1/2,-1}+H^2_{-1/2,0}\Big)+4\Big(H^2_{1/2,t}+H^2_{-1/2,t}\Big)\Big] , \\
	\frac{d\Gamma^{\lambda_l=-1/2}}{dq^2}&=&\frac{8}{3}\Big(H^2_{1/2,1}+H^2_{1/2,0}+H^2_{-1/2,-1}+H^2_{-1/2,0}\Big).
\end{eqnarray}

Having the form factors in hand, we can obtain the predictions of the total branching fractions, the averaged forward-backward asymmetry $\langle A_{FB}\rangle$, the averaged final hadron polarization $\langle P_{B}\rangle$ and the averaged lepton polarization $\langle P_{l}\rangle$. The numerical results of the relevant observables in the semi-leptonic decays  $\Lambda_b \rightarrow p(N^*)l^-\nu$ are presented  in Table.\ref{observable:proton} and Table.\ref{observable:Nstart} respectively. Similar to the prediction on the form factors, we list the results from  the Ioffe current and the tensor current as well as three different models of $\Lambda_b$ baryon for comparison. The lepton masses used in this work are taken from PDG\cite{PDG}, the central value of the life times of $\Lambda_b$ and the CKM matrix element $|V_{ub}|$ are adopted as $\tau_{\Lambda_b}=1.470$ps and   $|V_{ub}|=3.82\times 10^{-3}$ respectively. From the Table.\ref{observable:proton}, we can see that these physical observables for the semi-leptonic decay  $\Lambda_b \rightarrow pl^-\nu$ from two different interpolating currents and three LCDA models of $\Lambda_b$ baryon are consistent with each other. Our results are also consistent with the predictions from the relativistic quark model \cite{Faustov:2016pal} and the lattice simulation\cite{Dutta:2015ueb}, which is under expectation since the form factors are consistent as mentioned before.  Our estimation of the branch fraction of $Br({\Lambda_b \rightarrow p\mu^-\bar{\nu_{\mu}}})$ also agrees with the result reported by the LHCb Collaboration\cite{LHCb:2015eia}, and this channel provides a good  platform to determine the CKM matrix element $|V_{ub}|$ if the uncertainties can be reduced.  

 As for the physical observables of  the semi-leptonic decay $\Lambda_b \rightarrow N^*l\bar{\nu}$,  we can see from the Table \ref{observable:Nstart} that the results from the two different interpolating currents have large discrepancy.
The branching ratio  of $\Lambda_b \rightarrow N^*l\bar{\nu}$ decay from the tensor current is about seven times larger than that from the Ioffe current under the QCDSR and the Exponential models for the LCDA of $\Lambda_b-$baryon, and the ratio is even bigger when the Free-parton model is taken into account. For the averaged forward-backward asymmetry $\langle A_{FB}\rangle$ and the averaged final hadron polarization $\langle P_{B} \rangle$, the predictions from the Ioffe current  also significantly deviate from the results from the tensor current, which is mainly due to the smaller value of the $\Lambda_b \rightarrow N^*$ transition  form factors and the stronger LCDA-model dependency when the Ioffe current is adopted , see Table \ref{form factors at q0:Nstart}.  
Therefore, the physical observables of the semi-leptonic decay $\Lambda_b \rightarrow N^*l\bar{\nu}$  is very useful to figure out the properties of the baryon, such as the interpolating current of light-baryons, and the LCDAs of heavy-baryons can be better determined once   the physical observables  of heavy baryon semi-leptoinc decays are precisely measured. 

\section{Summary}
We have calculated the form factors of $\Lambda_b \rightarrow p, N^*(1535)$ transition within the framework of LCSR with the LCDAs of $\Lambda_b$-baryon, and utilized the results to evaluate the experimental observables such as the branching ratios, the forward-backward asymmetries and final state polarizations of the semileptonic decays $\Lambda_b \rightarrow p, N^*(1535)\ell \nu$. Since the interpolating current of the baryon is not unique, we employed three kinds of the current, namely the Ioffe current, the tensor current and the leading  power current for a comparison. Following a standard procedure of the calculation of heavy-to-light form factors by using LCSR approach, we can arrive at the sum rules of the $\Lambda_b \rightarrow p, N^*(1535)$ transition form factors. In the hadronic representation of the correlation function, we have isolated both proton state and the $N^*(1535)$ state so that the  $\Lambda_b \rightarrow p, N^*(1535)$ form factors can be evaluated simultaneously. The LCDAs of $\Lambda_b$-baryon are not well determined so far, thus we employed five different models, i.e, the QCDSR model, the exponential model, the free-parton model and the Gagenbauer-(I,II) models in order to find out the most suitable models.

Taking advantage of the leading power current, we can obtain the unique  $\Lambda_b \to p$ form factor at SCET limit, where only the LCDA $\psi_4$ of $\Lambda_b$-baryon appears in the sum rules. If we do not take the SCET limit, the numerical results will be only sightly changed. However, when the Ioffe current or tensor current are employed, we can obtain  much smaller values of the $\Lambda_b \rightarrow p$ form factors, and the results from the two interpolating currents are consistent with each other after including   the contribution of the negative-parity $N^*(1535)$ baryon in the hadronic dispersion relation. Since the correlation function defined  in terms of these two current contains some power suppressed terms, the numerical results indicate that the power suppressed contributions play a very important role in the sum rules of the $\Lambda_b \rightarrow p$ form factors. In  the numerical calculation, we have employed five different models for the LCDAs of $\Lambda_b$-baryon, and found that the results obtained from the models based on Gegenbauer polynomial expansion are too large and the results from the other three  models are  consistent with each other. Therefore, the  Gegenbauer expansion of the $\Lambda_b$ LCDAs is not appropriate  in the study on the $\Lambda_b$-baryon decays. Our predictions on the form factors for the $\Lambda_b \rightarrow p$ transition are consistent with the results of other theoretical works, especially  those from the QCD-inspired approaches such as light-hadron LCSR and Lattice QCD simulations.  As for the form factors of the $\Lambda_b \rightarrow N^*(1535)$ transition, the results are very sensitive to the choice of the interpolating currents and the uncertainties are very large. We only hope our predictions  will provide some useful information on the parameters of the $N^\ast$-LCDAs once the experimental data is available.

We further obtained the predictions of the total branching fractions, the averaged forward-backward asymmetry $\langle A_{FB}\rangle$, the averaged final hadron polarization $\langle P_{B}\rangle$ and the averaged lepton polarization $\langle P_{l}\rangle$.  Our results are  consistent with the predictions from the relativistic quark model\cite{Faustov:2016pal} and the lattice simulation\cite{Dutta:2015ueb}, and the prediction on the the branch fraction of $Br({\Lambda_b \rightarrow p\mu^-\bar{\nu_{\mu}}})$ also agrees with the measurement by the LHCb\cite{LHCb:2015eia}, and this channel provides a good  platform to determine the CKM matrix element $|V_{ub}|$. It also should be mentioned that the errors of our results are sizable, mainly due to  the large uncertainties of the parameters in the LCDAs of the $\Lambda_b$-baryon. Therefore, more studies on the $\Lambda_b$ baryon decays are required to constrain the parameters of the LCDAs of $\Lambda_b$ -baryon. Moreover, we only performed a tree-level calculation, and the QCD corrections to the hard kernel in the partonic expression of the correlation function are needed to increase the accuracy.  In the literatures\cite{Wang:2015ndk}, the QCD corrections to the leading power form factors of $\Lambda_b \to \Lambda$ have been calculated, the method can be directly generalized to the $\Lambda_b \to p$ transition. The power suppressed contributions have been shown to be important, and a more careful treatment on the power corrections is of great importance.   For example,  since  the LCDAs of $\Lambda_b$-baryon are defined in terms of the large component of the heavy quark field in HQET,  we have not included the power suppressed contributions from the heavy quark expansion, which will lead to an important kind of power correction. The above mentions problems will be considered in the future work. 

\section*{Acknowledgement}
We thank Yu-Ming Wang, Jia-Jie Han and Hua-Yu Jiang for useful discussions and valuable suggestions. 
This work was supported in part by the National Natural Science Foundation of China under
Grant Nos.~11975112,~12175218.
Y.L.S also acknowledges the Natural Science Foundation of Shandong province with Grant No. ZR2020MA093.

\appendix

\section{Correlation function in the LCSR for $\Lambda_b\rightarrow p , N^*(1535)$ form factor}\label{sec:AppendixA}
The invariant amplitudes $\Pi^i_j(p^2,q^2)$ for the correlation function  with the vector transition current $j_{\mu,V}$ in Eq.(11) are given in form of Eq.(20), where the  transformed coefficient functions $w^{(i)}_{jn}(s,q^2)$ with $i= {\rm  Ioffe, Tensor, LP}$ interpolating currents and $j=1 ,..., 6$(the invariant ampitude), $n=1,2$ (the power of the denominator $D$) are listed below:
\begin{itemize}
\item The Ioffe-type current
\begin{eqnarray}
	w^{\rm Io}_{11}&&=4f^{(1)}_{\Lambda_b}m_{\Lambda_b}\sigma\bar{\sigma}\psi^s_3\ ,\ \ \ \ \ 
		w^{\rm Io}_{12}=0,\,\,\,\,\,\,
	w^{\rm Io}_{21}=w^{\rm Io}_{61}=-\frac{2f^{(2)}_{\Lambda_b}}{\bar{\sigma}m^2_{\Lambda_b}}\tilde\psi_2 , \nonumber \\
w^{\rm Io}_{22}	&&=w^{\rm Io}_{62}=2f^{(2)}_{\Lambda_b}\frac{2m^2_{\Lambda_b}\sigma+q^2-m^2_{\Lambda_b}-s}{m^2_{\Lambda_b}}\tilde\psi_2-\frac{2f^{(2)}_{\Lambda_b}}{m_{\Lambda_b}}\tilde\Psi^{-+}_{12} ,\nonumber\\
  w^{\rm Io}_{31} &&	=-2f^{(1)}_{\Lambda_b}m^2_{\Lambda_b}\sigma\bar{\sigma}\psi^s_3+f^{(2)}_{\Lambda_b}[\tilde\psi_4+m_{\Lambda_b}\sigma(\psi^{+-}_1+\psi^{+-}_2)]\nonumber  \nonumber \\
	&&-f^{(2)}_{\Lambda_b}\left[\frac{(2s-3m_{\Lambda_b}^2\sigma-2q^2+m_{\Lambda_b}^2)}{m_{\Lambda_b}^2\bar{\sigma}}\tilde\psi_2-\frac{1}{m_{\Lambda_b}\bar{\sigma}}\tilde\Psi^{-+}_{12}\right],\nonumber \\	
w^{\rm Io}_{32}&&	=f^{(2)}_{\Lambda_b}\Big[\frac{(2m^2_{\Lambda_b}\sigma+q^2-m^2_{\Lambda_b}-s)(s-m_{\Lambda_b}^2\sigma-q^2)}{m^2_{\Lambda_b}}\tilde\psi_2\nonumber -\frac{s-q^2-m_{\Lambda_b}^2\sigma}{m_{\Lambda_b}}\tilde\Psi^{-+}_{12}\Big] , \nonumber
\\
	w^{\rm Io}_{41}&&=2f^{(1)}_{\Lambda_b}m_{\Lambda_b}\sigma\psi^s_3-\frac{f^{(2)}_{\Lambda}}{m_{\Lambda_b}\bar{\sigma}}\tilde\psi_2,
	\,\,w^{Io}_{42}=f^{(2)}_{\Lambda_b}\Big[\frac{2m^2_{\Lambda_b}\sigma+q^2-m^2_{\Lambda_b}-s}{m_{\Lambda_b}}\tilde\psi_2
	-\tilde\Psi^{-+}_{12}\Big] , \nonumber
\\
	w^{\rm Io}_{51}&&=-4f^{(1)}_{\Lambda_b}m_{\Lambda_b}\sigma^2\psi^s_3+\frac{2f^{(2)}_{\Lambda}}{m_{\Lambda_b}\bar{\sigma}}\tilde\psi_2\nonumber ,\,\,\,
	w^{Io}_{52}=2f^{(2)}_{\Lambda_b}\Big[-\frac{(2m^2_{\Lambda_b}\sigma+q^2-m^2_{\Lambda_b}-s)}{m_{\Lambda_b}}\tilde\psi_2
	+\tilde\Psi^{-+}_{12}\Big] , \nonumber
\end{eqnarray}
where 
$\tilde\Psi^{-+}_{12}=\tilde\psi^{-+}_1-\tilde\psi^{+-}_1+\tilde\psi^{-+}_2-\tilde\psi^{+-}_2$ and the functions $\tilde\psi(\omega,u)$ are defined as:
\begin{eqnarray}
	\tilde\psi(\omega, u)=\int^{\omega}_0 d\tau \ \tau\psi(\tau,u) ,
\end{eqnarray}
originating from the partial integral in the variable $\omega$ in Eq.(20)


\item The Tensor-type current
\begin{eqnarray}
	w^{\rm Te}_{11}&&=-2f^{(1)}_{\Lambda_b}m_{\Lambda_b}\sigma\Big[\bar{\sigma}(\psi_3^{+-}+2\psi_3^{-+})+\frac{2}{m_{\Lambda_b}}(\psi^2_{\perp,3}-\psi^1_{\perp,3})\Big],	\nonumber 	
	w^{\rm Te}_{12}=8f^{(1)}_{\Lambda_b}\bar{\sigma}(\tilde\psi^2_{\perp,Y}-\tilde\psi^1_{\perp,Y}),\nonumber
\\	w^{\rm Te}_{21}&&=w^{\rm Te}_{61}=0,\ \ \ \	
	w^{\rm Te}_{22}=w^{\rm Te}_{62}=\frac{2f^{(1)}_{\Lambda_b}}{m_{\Lambda_b}}(\tilde\psi^{+-}_{3}-\tilde\psi^{-+}_{3}) , \nonumber
\\
	w^{\rm Te}_{31}&&=-f^{(1)}_{\Lambda_b}m_{\Lambda_b}\sigma\Big[m_{\Lambda_b}\bar{\sigma}(\psi_3^{+-}+2\psi_3^{-+})+2(\psi^2_{\perp,3}-\psi^1_{\perp,3})\Big]-\frac{f^{(1)}_{\Lambda_b}}{m_{\Lambda_b}\bar{\sigma}}(\tilde\psi^{+-}_{3}-\tilde\psi^{-+}_{3}),	\nonumber \\	
	w^{\rm Te}_{32}&&=-f^{(1)}_{\Lambda_b}\Big[\frac{s-q^2-m^2_{\Lambda_b}\sigma}{m_{\Lambda_b}}(\tilde\psi^{+-}_{3}-\tilde\psi^{-+}_{3})-4m_{\Lambda_b}\bar{\sigma}(\tilde\psi^2_{\perp,Y}-\tilde\psi^1_{\perp,Y})\Big] , \nonumber
\\
	w^{\rm Te}_{41}&&=-f^{(1)}_{\Lambda_b}m_{\Lambda_b}\sigma(\psi_3^{+-}+2\psi_3^{-+}),	\nonumber 	
	w^{Te}_{42}=f^{(1)}_{\Lambda_b}\Big[(\tilde\psi^{+-}_{3}-\tilde\psi^{-+}_{3})+4\sigma(\tilde\psi^2_{\perp,Y}-\tilde\psi^1_{\perp,Y})\Big],\nonumber
\\
	w^{\rm Te}_{51}&&=2f^{(1)}_{\Lambda_b}m_{\Lambda_b}\sigma\Big[\sigma(\psi_3^{+-}+2\psi_3^{-+})-\frac{2}{m_{\Lambda_b}}(\psi^2_{\perp,3}-\psi^1_{\perp,3})\Big],	\nonumber 	\\
	w^{Te}_{52}&&=-2f^{(1)}_{\Lambda_b}\Big[(\tilde\psi^{+-}_{3}-\tilde\psi^{-+}_{3})+4\sigma(\tilde\psi^2_{\perp,Y}-\tilde\psi^1_{\perp,Y})\Big].\nonumber
\end{eqnarray}
\item The LP-type current
\begin{eqnarray}
	w^{LP}_{11}&&=-2f^{(2)}_{\Lambda_b}m_{\Lambda_b}\sigma\bar{\sigma}\psi_4	,\, \, \ \ \
	w^{LP}_{i2}(i=1-6)=0 ,\,\,
	w^{LP}_{21}=w^{LP}_{61}=0 , \nonumber\\
	w^{LP}_{31}&&=f^{(2)}_{\Lambda_b}m^2_{\Lambda_b}\sigma \bar{\sigma}\psi_4+f^{(1)}_{\Lambda_b}m_{\Lambda_b}\sigma(\psi^2_{\perp, 3}-\psi^1_{\perp, 3}+2(\psi^2_{\perp, Y}-\psi^1_{\perp, Y})) , \ \ \ \\
	w^{LP}_{41}&&=-f^{(2)}_{\Lambda_b}m_{\Lambda_b}\sigma\psi_4	,\,\, \ \ \
    w^{LP}_{51}=2f^{(2)}_{\Lambda_b}m_{\Lambda_b}\sigma^2\psi_4	.\ \ \ \ \
	\nonumber
\end{eqnarray}
\end{itemize}
The transformed coefficient functions $\tilde \omega^{(i)}_{jn}(s, q^2)$ for the correlation function with axial vector transition current $j_{\mu,A}$ can be obtained from $\omega^{(i)}_{jn}(s,q^2)$ in the above expressions.

\section{The LCDA models of $\Lambda_b$ baryon}
Light-cone distribution amplitudes(LCDA) of the $\Lambda_b$ baryon are the fundamental ingredients for the LCSR of the $\Lambda_b \rightarrow p(N^*)$ form factors, but they have attracted less attention compared to the $B$-meson. So far, only few works\cite{Ball:2008fw,Ali:2012pn,Bell:2013tfa} concerned about the LCDAs of the $\Lambda_b$ baryon, where different parameterization for the LCDAs of $\Lambda_b$ baryon have been proposed. In the present paper, we consider the following five different models:
\begin{itemize}
\item QCDSR model\cite{Ball:2008fw} 
\begin{eqnarray}
   \psi_2(\omega, u)&=&\frac{15}{2\mathcal{N}}\omega^2 u(1-u)\int^{s^{\Lambda_b}_0}_{\omega/2}ds\ e^{-s/\tau} (s-\omega/2), \nonumber \\ 
   \psi^{+-}_{3}(\omega,u)&=&\frac{15}{\mathcal{N}}\omega u\int^{s^{\Lambda_b}_0}_{\omega/2}ds\ e^{-s/\tau}(s-\omega/2)^2 , \nonumber \\
   \psi^{-+}_{3}(\omega,u)&=&\frac{15}{\mathcal{N}}\omega (1-u)\int^{s^{\Lambda_b}_0}_{\omega/2}ds\ e^{-s/\tau}(s-\omega/2)^2 , \nonumber \\
   \psi_4(\omega,u)&=&\frac{5}{\mathcal{N}}\int^{s^{\Lambda_b}_0}_{\omega/2}ds\ e^{-s/\tau}(s-\omega/2)^3,
\end{eqnarray}
with 
$\mathcal{N}=\int^{s^{\Lambda_b}_0}_0ds\ s^5e^{-s/\tau}$. The $\tau$ is the Borel parameter which is taken to be in the interval $0.4<\tau<0.8$ GeV and ${s^{\Lambda_b}_0}=1.2$ GeV is the continuum threshold.
\item Gegenbauer-1 model\cite{Ball:2008fw}
\begin{eqnarray}
 \psi_{2}(\omega, u)&=&\omega^2u(1-u)[\frac{1}{\epsilon^4_0}e^{-\omega/\epsilon_0}+a_2C^{3/2}_{2}(2u-1)\frac{1}{\epsilon^4_1}e^{-\omega/\epsilon_1}], \nonumber \\
 \psi^{+-}_{3}(\omega, u)&=&\frac{2}{\epsilon^3_3}\omega ue^{-\omega/\epsilon_3}, \nonumber \\
  \psi^{-+}_{3}(\omega, u)&=&\frac{2}{\epsilon^3_3}\omega (1-u)e^{-\omega/\epsilon_3}, \nonumber \\
 \psi_4(\omega,u)&=&\frac{5}{\mathcal{N}}\int^{s^{\Lambda_b}_0}_{\omega/2}ds\ e^{-s/\tau}(s-\omega/2)^3,
\end{eqnarray}
with $C^{3/2}_2$ is Gegenbauer polynomial, and $\epsilon_0=200^{+130}_{-60}$ Mev, $\epsilon_1=650^{+650}_{-300}$ Mev, $a_2=0.333^{+0.250}_{-0.333}$ and $\epsilon_3=230\pm 60$ MeV. 
\item Gegenbauer-2 model\cite{Ali:2012pn}\\
\begin{eqnarray}
    \psi_{2}(\omega, u)&=&\omega^2 u(1-u)\sum^2_{n=0}\frac{a_n}{\epsilon^{4}_{n}}\frac{C^{3/2}_{n}(2u-1)}{|C^{3/2}_n|^2}e^{-\omega/\epsilon_n}, \nonumber \\
    \psi^{\sigma,s}_{3}(\omega, u)&=&\frac{\omega}{2} \sum^{2}_{n=0}\frac{a_n}{\epsilon^3_n}\frac{C^{1/2}_n (2u-1)}{|C^{1/2}_n|^2}e^{-\omega/\epsilon_n},\nonumber \\
    \psi_{4}(\omega, u)&=&\sum^2_{n=0}\frac{a_n}{\epsilon^2_n}\frac{C^{1/2}_n (2u-1)}{|C^{1/2}_n|^2}e^{-\omega/\epsilon_n},
\end{eqnarray}
with $|C^{1/2}_{0}|^2=|C^{3/2}_{0}|^2=1,\ |C^{1/2}_1|^2=1/3,\ |C^{3/2}_1|^2=3,\ |C^{1/2}_2|^2=1/5$ and $|C^{3/2}_2|^2=6$. The parameters in the above are collected in Tab.\ref{tab:parameter} and $A=0.5\pm0.1$. The twist-3 LCDAs $\psi^{+-}_3, \psi^{-+}_3$ are given by the combination of $\psi^{\sigma,s}_3$
\begin{eqnarray}
    \psi^{+-}_3(\omega, u)&=&2\psi^{s}_3(\omega, u)+2\psi^{\sigma}_3(\omega, u) ,\nonumber \\
    \psi^{-+}_3(\omega, u)&=&2\psi^{s}_3(\omega, u)-2\psi^{\sigma}_3(\omega, u),
\end{eqnarray}
\begin{table}[h]
    \centering
    \caption{Parameters for the LCDA of $\Lambda_b$ baryon}
    \begin{tabular}{c |c| c c c c c c}
    \hline
    \hline
    \multirow{5}{*}{$\Lambda_b$}&twist&$a_0$&$a_1$&$a_2$&$\epsilon_0$[Gev]&$\epsilon_1$[Gev]&$\epsilon_2$[Gev]\\\cline{2-8}
    \hline
    &2&1&-&$\frac{6.4(1-A)}{1.44-A}$&$\frac{2.0-1.4A}{6.7-A}$&-&$\frac{0.32(1-A)}{0.83-A}$\\
    &3s&1&-&$\frac{0.12A-0.04}{A+0.4}$&$\frac{0.56A+0.21}{A+1.6}$&-&$\frac{0.09-0.25A}{1.41-A}$\\
    &$3\sigma$&-&1&-&-&$\frac{0.35A+0.08}{A+0.2}$&-\\
    &4&1&-&$\frac{0.07A-0.12}{1.34-A}$&$\frac{0.87-0.65A}{2-A}$&-&$\frac{9.3-5.5A}{30-A}$\\
    \hline
    \hline
    \end{tabular}
    \label{tab:parameter}
\end{table}
\item Exponential model\cite{Bell:2013tfa}
\begin{eqnarray}
    \psi_2(\omega,u)&=&\frac{\omega^2 u(1-u)}{\omega^4_0}e^{-\omega/\omega_0}, \nonumber \\
    \psi^{+-}_3(\omega, u)&=&\frac{2\omega u}{\omega^3_0}e^{-\omega/\omega_0} ,\nonumber \\
    \psi^{-+}_3(\omega, u)&=&\frac{2\omega (1-u)}{\omega^3_0}e^{-\omega/\omega_0} ,\nonumber \\
        \psi_4(\omega, u)&=&\frac{1}{\omega^2_0}e^{-\omega/\omega_0},
\end{eqnarray}
where $\omega_0=0.4\pm 0.1$ GeV measures the average of the two light quarks inside the $\Lambda_b$ baryon.
\item Free parton model\cite{Bell:2013tfa}
\begin{eqnarray}
    \psi_2(\omega, u)&=&\frac{15\omega^2 u(1-u)(2\bar{\Lambda}-\omega)}{4\bar{\Lambda}^5}\theta(2\bar{\Lambda}-\omega), \nonumber \\
    \psi^{+-}_3(\omega, u)&=&\frac{15\omega u (2\bar{\Lambda}-\omega)^2}{4\bar{\Lambda}^5}\theta(2\bar{\Lambda}-\omega), \nonumber \\
    \psi^{-+}_3(\omega, u)&=&\frac{15\omega (1-u) (2\bar{\Lambda}-\omega)^2}{4\bar{\Lambda}^5}\theta(2\bar{\Lambda}-\omega), \nonumber \\
    \psi_4(\omega, u)&=&\frac{5(2\bar{\Lambda}-\omega)^3}{8\bar{\Lambda}^5}\theta(2\bar{\Lambda}-\omega) ,
\end{eqnarray}
where $\theta(2\bar{\Lambda}-\omega)$ is the step-function, and $\bar{\Lambda}=m_{\Lambda_b}-m_b\approx 0.8\pm0.2$ GeV.
\end{itemize}

The LCDAs of the NLO term off the light-cone are given by :
\begin{eqnarray}
  \psi^{+-}_{\perp, 1}(\omega, u)&=&  \psi^{-+}_{\perp, 2}=\psi^{(1)}_{\perp, 3}=\psi^{(2)}_{\perp, 3}=\frac{\omega^2 u(1-u)}{\omega_0^3} e^{-\omega/\omega_0} ,\nonumber \\
  \psi^{-+}_{\perp, 1}(\omega, u)&=&\frac{\omega u}{\omega_0^2} e^{-\omega/\omega_0} , \nonumber \\
  \psi^{+-}_{\perp, 2}(\omega, u)&=&\frac{\omega (1-u)}{\omega_0^2} e^{-\omega/\omega_0} , \nonumber \\
  \psi^{(1)}_{\perp, Y}(\omega, u)&=&\frac{\omega u(\omega_0-\omega(1-u))}{2\omega_0^3}e^{-\omega/\omega_0} ,\nonumber \\
  \psi^{(2)}_{\perp, Y}(\omega, u)&=&\frac{\omega (1-u)(\omega_0-\omega u)}{2\omega_0^3}e^{-\omega/\omega_0}.
\end{eqnarray}
where $\omega_0=0.4\pm0.1$ GeV.


\end{document}